\documentclass{aastex}
\usepackage{spr-astr-addons}
\usepackage{url}
\usepackage{amsmath}
\usepackage{amssymb}
\allowdisplaybreaks

\RequirePackage{color}

\newcommand{\emaila}{authors@email.com}

\shorttitle{On the conservation of the Jacobi integral in the post-Newtonian CR3BP}
\shortauthors{F. L. Dubeibe, F. D. Lora-Clavijo \& G. A. Gonz\'alez}

\begin{document}

\title{On the conservation of the Jacobi integral in the post - Newtonian circular restricted three-body problem}

\author{F. L. Dubeibe\altaffilmark{1,2}} 
\affil{fldubeibem@unal.edu.co}
\and 
\author{F. D. Lora-Clavijo\altaffilmark{2}}
\affil{fadulora@uis.edu.co}
\and
\author{Guillermo A. Gonz\'alez\altaffilmark{2}}
\affil{guillermo.gonzalez@saber.uis.edu.co}
\email{\emaila fldubeibem@unal.edu.co}
\altaffiltext{1}{Facultad de Ciencias Humanas y de la Educaci\'on, Universidad de los Llanos, Villavicencio, Colombia.}
\altaffiltext{2}{Grupo de Investigaci\'on en Relatividad y Gravitaci\'on, Escuela de F\'isica, Universidad Industrial de Santander, A.A. 678, Bucaramanga, Colombia.}

\begin{abstract}
In the present paper, using the first-order approximation of the $n$-body Lagrangian (derived on the basis of the post-Newtonian gravitational theory of Einstein, Infeld, and Hoffman), we explicitly write down the equations of motion for the planar circular restricted three-body problem. Additionally, with some simplified assumptions, we obtain two formulas for estimating the values of the mass-distance and velocity-speed of light ratios appropriate for a given post-Newtonian approximation.  We show that the formulas derived in the present study, lead to a good numerical conservation of the Jacobi constant and allow for an approximate equivalence between the Lagrangian and Hamiltonian approaches at the same post-Newtonian order. Accordingly, the dynamics of the system is analyzed in terms of the Poincaré sections method and Lyapunov exponents, finding that for specific values of the Jacobi constant the dynamics can be either chaotic or regular. Our results suggest that the chaoticity of the post-Newtonian system is slightly in- creased in comparison with its Newtonian counterpart.
\end{abstract}


\section{Introduction}

The circular restricted three-body problem (CR3BP) describes the gravitational interaction between three masses, in which two primaries $m_1$ and $m_2$ are moving in circular orbits around their common center of mass, while a third body  $m$ whose mass is very small compared to the others, moves in the plane of the primaries. Many studies have been carried out for the Newtonian CR3BP \citep{Marchal2012}. In particular, the stability of the Lagrangian points has attracted the attention of many researchers \citep{Celletti1990,Bombardelli2011,Salazar2012}, due to the possible applications to different astronomical systems \citep{Musielak2014}: Earth-Moon system with a spacecraft, Sun-Earth-Moon system, Sun-Jupiter system with an asteroid, Binary stellar systems with exoplanets, among others. Concerning dynamical studies, Gonczi and Froeschl\'e, by using the Lyapunov characteristic numbers \citep{Gonczi1981} and surfaces of section \citep{Froeschle1970}, respectively, have shown that the Newtonian CR3BP is chaotic.

To account for relativistic effects on motions of astronomical bodies, Krefetz extended the investigation of the CR3BP to General relativity \citep{Krefetz1967}. Instead of solving the full Einstein equations, he used the Einstein-Infeld-Hoffmann (EIH) theory \citep{Einstein1938} at the first post-Newtonian approximation (1-PN), to calculate the equations of motion in a synodic frame of reference. An alternative derivation of these equations for the general and restricted three-body problem was performed by \citet{Brumberg1972}, who introduced a new parameter $n$ denoting the mean motion of the primaries. Many years later, Maindl and Dvorak re-derived the equations of motion of the test particle $m$ and applied its results to the computation of the advance of Mercury perihelion \citep{Maindl1994}. It is important to note that despite the fact that all the three different approaches used the EIH formalism up to 1-PN order, do not lead to the same set of equations, mainly due to the introduction of different relativistic parameters. The question of the equivalence (or not) of the Lagrangian and Hamiltonian approaches at the same post- Newtonian (PN) order is also a topic of current interest. The equivalence between both approaches was shown in \cite{Damour2001}, \cite{Damour2002}, \cite{Andrade2001} and \cite{Levi2014}. However, different claims exist in \cite{WuPRD}, \cite{WuMNRAS}, \cite{Wang2015}, \cite{Chen2016} and \cite{Huang2016}, stating that such differences are owed to the truncation of higher-order PN terms.

The study of the dynamics in the post-Newtonian restricted three body problem is very recent. In \citeyear{Huang2014} Huang and Wu studied the effects on the dynamics of the distance between the primaries by applying scaling transformations to distance and time. To do so, the authors used the equations derived by \citet{Maindl1994}, including the modified angular speed outlined by \citet{Contopoulos1976} for the 1-PN circular two-body problem. As we will show in the present paper, the use of the Maindl's equations in canonical units with the choice $c=1$ is not adequate to perform dynamical studies, due to the non-conservation of the Jacobi constant. Taking into account the previous arguments, in this paper we find an alternative system of equations of motion for the planar CR3BP using the EIH formalism up to 1-PN order. Two formulas for the determination of appropriate values of the mass-distance and velocity-speed of light ratios in a general $n$-th order post-Newtonian approximation are derived, suggesting that in canonical units a suitable value of $c$ should be of the order $10^{4}$. Under these conditions, we show that the Jacobi constant is conserved to a good numerical approximation and additionally, the 1PN Lagrangian and Hamiltonian approaches are equivalent.  

The paper is organized as follows: in section \ref{sec:1}, we consider the Lagrangian for $n$-bodies in the first-order post-Newtonian approximation and derive the equations of motion for a test particle in the planar circular restricted three-body problem. The region of validity of the PN approximations is discussed in section \ref{sec:2}. The comparison of the Lagrangian and Hamiltonian approaches is carried out in section \ref{sec:4}. In section \ref{sec:3}, we compare the conservation of the Jacobi integral obtained with three different approaches: the one presented in this work, the one derived by Maindl and the used by Huang and Wu. Additionally, a dynamical analysis in terms of the Poincar\'e surfaces of section and Lyapunov exponents is performed. Finally, in section \ref{sec:5} we summarize our main conclusions. 

\section{Post-Newtonian equations of motion}
\label{sec:1}
The first-order post-Newtonian equations of motion for an $n$-body system interacting with each other gravitationally were first obtained by \cite{Einstein1938} and independently by \cite{Eddington1938}. The Lagrangian associated with these equations of motion reads as  \citep{Landau2013},
\begin{align}\label{eq1}
{\cal L}&=\sum_{i=1}^{n}\frac{m_{i} v_{i}^2}{2}\left(1+3\sum_{j\ne i}^{n} \frac{G m_{j}}{c^2 r_{i j}}\right)+\sum_{i=1}^{n}\frac{m_{i} v_{i}^4}{8 c^2}\nonumber\\
&-\sum_{i=1}^{n}\sum_{j\ne i}^{n}\frac{G m_{i} m_{j}}{4 c^2 r_{i j}}\left[7\mathbf{v}_{i}\cdot\mathbf{v}_{j}+(\mathbf{v}_{i}\cdot\mathbf{n}_{i j})(\mathbf{v}_{j}\cdot\mathbf{n}_{i j})\right]\nonumber\\
&-\sum_{i=1}^{n}\sum_{j\ne i}^{n}\sum_{k\ne i}^{n}\frac{G^2 m_{i} m_{j} m_{k}}{2 c^2 r_{i j} r_{i k}} +\sum_{i=1}^{n}\sum_{j\ne i}^{n}\frac{G m_{i} m_{j}}{2 r_{i j}},\nonumber\\
&+ O(c^{-4}),
\end{align}
with $\mathbf{v}_{i}=d\mathbf{r}_{i}/dt$, $r_{i j}=|\mathbf{r}_{i} - \mathbf{r}_{j} |$ and  $\mathbf{n}_{i j}=(\mathbf{r}_{i} - \mathbf{r}_{j})/|\mathbf{r}_{i} - \mathbf{r}_{j}|$.

When considering the Lagrangian of a single particle with mass $m$, at position $\mathbf{r}$, with velocity $\mathbf{u}$, in the gravitational field of two other bodies at positions $\mathbf{r}_{i}$ with respective velocities $\mathbf{v}_{i}$, the expression \eqref{eq1} can be rewritten as \citep{Contopoulos1967}
\begin{align}\label{eq2}
\nonumber{\cal L}&=  \frac{m \mathbf{u}^{2}}{2} + \frac{m\mathbf{u}^{4}}{8c^{2}}+ Gm\sum_{i=1}^{2}\frac{m_{i}}{|\mathbf{r} - \mathbf{r}_{i} |} - \frac{G^{2}m}{2c^{2}}\\ \nonumber
&\times\left( \sum_{i=1}^{2}\frac{m_{i}}{|\mathbf{r} - \mathbf{r}_{i} |} \right)^{2} 
- \frac{Gm}{2c^{2}}\sum_{i=1}^{2}\frac{m_{i}}{|\mathbf{r} - \mathbf{r}_{i} |}
\Bigg[ 7\left(\mathbf{u}\cdot\mathbf{v}_{i}\right) \\\nonumber
&+ \left(\mathbf{n}_{i}\cdot\mathbf{u}\right)\left(\mathbf{n}_{i}\cdot\mathbf{v}_{i}\right) -3\left(\mathbf{u}^{2} + \mathbf{v}_{i}^{2}\right) + 2G\sum_{j\ne i}^{2}\frac{m_{j}}{|\mathbf{r}_{i} - \mathbf{r}_{j} |} \Bigg]\\
 &+ O(c^{-4}),
\end{align}
with $\mathbf{n}_{i}=(\mathbf{r} - \mathbf{r}_{i})/|\mathbf{r} - \mathbf{r}_{i}|$.  

The planar circular restricted three-body problem (henceforth PCR3BP) is a special case of the Lagrangian \eqref{eq2}, in which the primaries $m_1$ and $m_2$ move in circular orbits around their common center of mass at a fixed distance $a$. Under these considerations and introducing a synodic frame of reference through the transformations, 
\begin{eqnarray}
\mathbf{r}&=&\{x\cos\omega t -y \sin\omega t, x\sin\omega t +y \cos\omega t\},\nonumber\\
\mathbf{u}&=&\{(\dot{x}-\omega y)\cos\omega t -(\dot{y}+\omega x) \sin\omega t,\nonumber\\
&&(\dot{y}+\omega x) \cos\omega t +(\dot{x}-\omega y)\sin\omega t\},\nonumber\\
\mathbf{r}_{i}&=&\{x_{i}\cos\omega t, x_{i}\sin\omega t\},\nonumber\\
\mathbf{v}_{i}&=&\{-\omega x_{i}\sin\omega t, \omega x_{i}\cos\omega t\},\nonumber
\end{eqnarray}
with $i=1,2$ and $\omega$ the angular velocity of the primaries, the Lagrangian (\ref{eq2}) of the third particle (cf. \cite{Maindl1994,Huang2014}) can be reduced to
\begin{eqnarray}\label{Lf}
\frac{\cal L}{m} &=& {\cal L_{N}} + \frac{1}{c^{2}} {\cal L_{R}},
\end{eqnarray}
with ${\cal L}_{N}$ the Newtonian Lagrangian and ${\cal L}_{R}$ the relativistic 1-PN contribution, {\it i.e.} 
\begin{eqnarray}
{\cal L}_{N}=\frac{1}{2}\Omega^2 + G \left(\frac{m_{1}}{d_{1}} + \frac{m_{2}}{d_{2}}\right),
\end{eqnarray}
\begin{align}
{\cal L}_{R}&=
-G^2\left[\frac{m_{1}m_{2}}{a}\left(\frac{1}{d_{1}} + \frac{1}{d_{2}}\right)
+\frac{1}{2}\left(\frac{m_{1}}{d_{1}} + \frac{m_{2}}{d_{2}}\right)^2\right]\nonumber\\ 
&+\frac{1}{8}\Omega^{4} +\frac{G}{2}\bigg\{3\left(\frac{m_{1}}{d_{1}} + \frac{m_{2}}{d_{2}}\right)\Omega^{2}
- 7\omega(\dot{y}+\omega x) \nonumber\\ 
&\times\bigg(\frac{m_{1}x_{1}}{d_{1}}+\frac{m_{2}x_{2}}{d_{2}}\bigg)+ 3\omega^{2}\left(\frac{m_{1}x_{1}^{2}}{d_{1}} + \frac{m_{2}x_{2}^2}{d_{2}}\right)\nonumber\\ 
&- \omega y \left( \dot{x} - \omega y \right)\bigg[\frac{m_{1}x_{1}\left(x-x_{1}\right)}{d_{1}^{3}} + \frac{m_{2}x_{2}\left(x-x_{2}\right)}{d_{2}^{3}}\bigg] \nonumber\\
&- \omega y^{2} ( \dot{y} + \omega x)\left(\frac{m_{1}x_{1}}{d_{1}^{3}} + \frac{m_{2}x_{2}}{d_{2}^{3}}\right)
\bigg\},
\end{align}
where $\Omega^2=U^{2} + 2 A \omega + R^{2}\omega^{2}, d_{1}^2 = \left(x - x_{1}\right)^{2} + y^{2}, d_{2}^2 = \left(x - x_{2}\right)^{2} + y^{2}, U^{2} = \dot{x}^{2} + \dot{y}^{2}, A = \dot{y}x-\dot{x}y$ and $R^{2} = x^{2} + y^{2}$.
Here, $x_{i}$ denotes the fixed position of the primary $m_{i}$ in the synodic frame of reference. The corresponding angular velocity up to the 1-PN-approximation was calculated by \cite{Contopoulos1976} and reads as 
\begin{equation}\label{eq:omega}
\omega^2=\frac{G(m_1+m_2)}{a^3}\left\{1+\frac{G\left[m_{1}m_{2}-3(m_1+m_2)^2\right]}{a c^2 (m_1+m_2)}\right\}.
\end{equation}

In analogy with the Newtonian case, the Post-Newtonian Jacobi integral (or the so called Jacobi constant) can be derived from the energy function $h$ \citep{Contopoulos2013}, that is, the energy function $h$ is generated by the Legendre transformation 
\begin{equation}
h=\sum_{i=1}^{2} p_{i} \dot{q}_{i} - {\cal L}(q_{i}, \dot{q}_{i}),
\end{equation}
with generalized momenta
\begin{equation}\label{eq:pi}
p_{i}=\frac{\partial {\cal L}}{\partial \dot{q}_{i}},
\end{equation}
thus, we may define the Jacobi constant as
\begin{equation}\label{eq:Jacg}
-J=h=\sum_{i=1}^{2} \dot{q}_{i} \frac{\partial {\cal L}}{\partial \dot{q}_{i}} - {\cal L}(q_{i}, \dot{q}_{i}).
\end{equation}

Taking into account that the Hamiltonian ${\cal H}$ is constructed in the same manner as $h$, but 
they are functions of different variables, the energy function can be written in terms of positions and momenta so that in our setting, the Jacobi constant and the Hamiltonian satisfy the relation 
 \begin{equation}\label{eq:hH}
 J=-\cal{H}.
 \end{equation}
Under this condition, the Jacobi constant becomes the only known conserved quantity $dJ/dt = \{J,{\cal H}\} = 0$, due to the absence of cyclic coordinates.\footnote{In the Newtonian limit $1/c^2\rightarrow 0$, $J=\omega L-E$, with $E$ the total energy and $L$ the angular momentum.}

Therefore, taking into account the equation for the angular velocity \eqref{eq:omega}, substituting \eqref{Lf} into \eqref{eq:Jacg} and keeping terms up to the order $1/c^2$, we get  
\begin{eqnarray}\label{Jacob}
J&=& G \bigg(\frac{m_1}{d_1}+\frac{m_2}{d_2}\bigg)+\frac{1}{2}\left(R^2 \omega_{0}^2-U^2\right)
+\frac{1}{c^2}\bigg\{\frac{R^4 \omega_{0}^4}{8}\nonumber\\
&-&\frac{3 U^4}{8}-\frac{R^2\omega_{0}^2 U^2}{4}-A U^2 \omega_{0}+\omega_{0}^2 \omega_{1} R^{2} -\frac{A^2 \omega_{0}^2}{2}\nonumber\\
&-&\frac{3}{2} G \left(\frac{m_{1}}{d_{1}}+\frac{m_{2}}{d_{2}}\right)\left(U^2-R^2 \omega_{0}^2\right)
+\frac{3}{2} G \omega_{0}^2\nonumber\\
&\times&\left(\frac{m_{1} x_{1}^2}{d_{1}}+\frac{m_{2} x_{2}^2}{d_{2}}\right)-\frac{7}{2} G x \omega_{0}^2 \left(\frac{m_{1} x_{1}}{d_{1}}+\frac{m_{2} x_{2}}{d_{2}}\right)\nonumber\\
&-&\frac{G y^2 \omega_{0}^2}{2} \left(\frac{m_{1} x_{1}^2}{d_{1}^3}+\frac{m_{2} x_{2}^2}{d_{2}^3}\right)-\frac{G^2}{2}  \left(\frac{m_{1}}{d_{1}}+\frac{m_{2}}{d_{2}}\right)^2\nonumber\\
&-&G^2 \frac{m_{1} m_{2}}{a}\left(\frac{1}{d_{1}}+\frac{1}{d_{2}}\right)\bigg\},
\end{eqnarray}
where $\omega_{1} = G \left[m_{1}m_{2} - 3 (m_1 + m_2)^2\right]/2a(m_1+m_2)$ and $\omega_{0}^{2} = G(m_1+m_2)/a^3$. Eq. \eqref{Jacob} coincides exactly with the expression obtained by \cite{Maindl1994}.

The equations of motion for the test particle in the PCR3BP can be calculated from the Lagrangian (\ref{Lf}), with the aid of the Euler-Lagrange equations,
\begin{equation}\label{eq:E-L}
\frac{d}{dt}\left(\frac{\partial {\cal L}}{\partial \dot{q}_{i}}\right)-\frac{\partial {\cal L}}{\partial q_{i}}=0.
\end{equation}
From Eqs. (\ref{Lf}-\ref{eq:omega}), \eqref{eq:E-L}, and neglecting terms higher than $1/c^{2}$, we get
a system of two equations of the form
\begin{eqnarray}
f(x,y,\dot{x},\dot{y})+\ddot{x} C_{1x}(x,y,\dot{x},\dot{y})+\ddot{y} C_{2x}(x,y,\dot{x},\dot{y})&=&0,\nonumber\\
g(x,y,\dot{x},\dot{y})+\ddot{x} C_{1y}(x,y,\dot{x},\dot{y})+\ddot{y} C_{2y}(x,y,\dot{x},\dot{y})&=&0,\nonumber
\end{eqnarray}
with $C_{2x}(x,y,\dot{x},\dot{y})=C_{1y}(x,y,\dot{x},\dot{y})$, whose solutions are
\begin{eqnarray}
\ddot{x}(C_{1y}^2-C_{1x}C_{2y})&=&C_{2y} f-C_{1y} g,\label{eq:xdd-gen}\\
\ddot{y}(C_{1y}^2-C_{1x}C_{2y})&=&C_{1x} g-C_{1y} f\label{eq:ydd-gen}.
\end{eqnarray}

Substituting the corresponding expressions for $C_{1x}$, $C_{1y}$, $C_{2y}$, $f$ and $g$, and again neglecting terms higher that $1/c^{2}$ in \eqref{eq:xdd-gen} and \eqref{eq:ydd-gen}, the resulting equations of motion can be explicitly written as
\begin{eqnarray}
\left[1 + \frac{2\Omega^2}{c^2} +  \frac{6 G}{c^2}\left(\frac{m_{1}}{d_{1}} + \frac{m_{2}}{d_{2}} \right)\right] \ddot{x} &=& \left(\ddot{x}_{c} + \frac{\ddot{x}_{r}}{c^2}   \right) \label{eq:xppf}\\ 
\left[1 + \frac{2\Omega^2}{c^2} +  \frac{6 G}{c^2}\left(\frac{m_{1}}{d_{1}} + \frac{m_{2}}{d_{2}} \right)\right] \ddot{y} &=& \left(\ddot{y}_{c} + \frac{\ddot{y}_{r}}{c^2}   \right)\label{eq:yppf}
\end{eqnarray}
where $\ddot{x}_{c}$, $\ddot{y}_{c}$, $\ddot{x}_{r}$ and  $\ddot{y}_{r}$ are given by the following expressions 
\begin{eqnarray}
\ddot{x}_{c} &=& x + 2\dot{y} - G \left[ \frac{m_{1}(x-x_{1})}{d_{1}^{3}}  +  \frac{m_{2}(x-x_{2})}{d_{2}^{3}} \right], \\ 
\ddot{y}_{c} &=& y - 2\dot{x} - G y \left(  \frac{m_{1}}{d_{1}^{3}}  +  \frac{m_{2}}{d_{2}^{3}} \right),
\end{eqnarray}
\begin{eqnarray}
\ddot{x}_{r} &=& 2 \left(x+ \dot{y}\right) \omega_{1} +2(x+2\dot{y})\left[ \left(\dot{x} -y \right)^{2} + \left(x + \dot{y} 
\right)^{2}\right]  \nonumber\\
&+& G^{2}\Bigg[  \frac{m_{1}^{2}(x-x_{1})}{d_{1}^{4}} + \frac{m_{2}^{2}(x-x_{2})}{d_{2}^{4}} + \frac{m_{1}m_{2}}{a} \nonumber\\ 
&\times& \left( \frac{x-x_{1}}{d_{1}^{3}} + \frac{x-x_{2}}{d_{2}^{3}} \right) + \frac{m_{1}m_{2}}{d_{1} d_{2}}\Bigg( \frac{x-x_{1}}{d_{1}^{2}} \nonumber\\
&+& \frac{x-x_{2}}{d_{2}^{2}} \Bigg) - 3 \left(  \frac{m_{1}}{d_{1}}  +  \frac{m_{2}}{d_{2}} \right)\Bigg(\frac{m_{1}^{2}(x-x_{1})}{d_{1}^{3}} \nonumber\\
&+& \frac{m_{2}^{2}(x-x_{2})}{d_{2}^{3}}\Bigg) \Bigg]  + G \Bigg[-\frac{7}{2} \left(\frac{m_{1}x_{1}}{d_{1}}+\frac{m_{2}x_{2}}{d_{2}} \right) \nonumber\\
&+& 5x \left( \frac{m_{1}(x-x_{1})x_{1}}{d_{1}^{3}} +\frac{m_{2}(x-x_{2})x_{2}}{d_{2}^{3}} \right) \nonumber\\
&-&  \frac{3}{2}\left( \frac{m_{1}(x^{3}-x_{1}^{3})}{d_{1}^{3}} +\frac{m_{2}(x^{3}-x_{2}^{3})}{d_{2}^{3}}  \right)\nonumber \\
&+& \frac{3}{2} \left( \frac{m_{1}(x-x_{1})x_{1}^{2}}{d_{1}^{5}} +\frac{m_{2}(x-x_{2})x_{2}^{2}}{d_{2}^{5}} \right)y^{2}\nonumber \\ 
&+& \left( \frac{m_{1}(x-x_{1})}{d_{1}^{3}} +\frac{m_{2}(x-x_{2})}{d_{2}^{3}} \right)\nonumber \\
&\times& \left( (\dot{x} -y)(\dot{x}+2 y) - \frac{3}{2} (x^{2}+2x\dot{y}+2\dot{y}^{2}) \right) \nonumber\\
&+& 7x \left(  \frac{m_{1}x_{1}}{d_{1}^{3}}  +  \frac{m_{2}x_{2}}{d_{2}^{3}} \right)\dot{y} - 4 \left(  \frac{m_{1}x_{1}^{2}}{d_{1}^{3}}  +  \frac{m_{2}x_{2}^{2}}{d_{2}^{3}} \right)\dot{y}  \nonumber\\
&+& 6 \left(  \frac{m_{1}}{d_{1}}  +  \frac{m_{2}}{d_{2}} \right)(x+2\dot{y}) + \left(  \frac{m_{1}}{d_{1}^{3}}  +  \frac{m_{2}}{d_{2}^{3}} \right)\nonumber \\ 
&\times& \left( x(\dot{x} - y)y -3x^{2}\dot{y} + 4(\dot{x}-y) y \dot{y}\right) \bigg ],
\end{eqnarray}
\begin{eqnarray}
\ddot{y}_{r} &=&2(y-\dot{x}) \omega_{1} - 2(2\dot{x} -y)\left( (\dot{x}-y)^{2} + (\dot{y}+x)^{2} \right)\nonumber\\
&+& G^{2} \bigg[\frac{m_{1}m_{2}y}{a}\bigg( \frac{1}{d_{1}^{3}}+\frac{1}{d_{2}^{3}} \bigg) - 2 \bigg( \frac{m_{1}}{d_{1}^{3}}+\frac{m_{2}}{d_{2}^{3}} \bigg)\nonumber\\ 
&\times&\bigg( \frac{m_{1}}{d_{1}}+\frac{m_{2}}{d_{2}} \bigg) y\bigg] + G\bigg[-6 \bigg( \frac{m_{1}}{d_{1}}+\frac{m_{2}}{d_{2}} \bigg)\nonumber\\ 
&\times&(2\dot{x} -y) + \frac{1}{2} \bigg( \frac{m_{1}x_{1}^{2}}{d_{1}^{3}}+\frac{m_{2}x_{2}^{2}}{d_{2}^{3}} \bigg)(8\dot{x} -5y)\nonumber\\ 
&+&  \frac{3}{2} \bigg( \frac{m_{1}x_{1}^{2}}{d_{1}^{5}}+\frac{m_{2}x_{2}^{2}}{d_{2}^{5}} \bigg)y^{3} + \frac{7}{2} x\bigg( \frac{m_{1}x_{1}}{d_{1}^{3}}+\frac{m_{2}x_{2}}{d_{2}^{3}} \bigg)\nonumber\\ 
&\times& (y-2\dot{x} )  +\bigg(   \frac{m_{1}(x-x_{1})}{d_{1}^{3}} +\frac{m_{2}(x-x_{2})}{d_{2}^{3}} \bigg)\nonumber\\
&\times&(x(\dot{x}-y)+(4\dot{x}-y)\dot{y}) + \bigg( \frac{m_{1}}{d_{1}^{3}}+\frac{m_{2}}{d_{2}^{3}} \bigg)\nonumber\\
&\times&(x^{2} (3\dot{x}-2y)-xy\dot{y} + y(\dot{y}^2-3(\dot{x}-y)^{2})) \bigg].
\end{eqnarray}

Some important points should be noted in relation to the above equations: \\{\it (i)} In the non-relativistic limit $1/c^2\to 0$, the equations of motion reduce to their Newtonian expressions. \\{\it (ii)} The equations of motion are consistent with the 1-PN-approximation. \\{\it (iii)} Equations \eqref{eq:xppf} and \eqref{eq:yppf} do not coincide with the ones derived by \cite{Maindl1994} and used by \cite{Huang2014}. \\{\it (iv)} Due to the approximations carried out in Eqs. \eqref{eq:xppf} and \eqref{eq:yppf}, the Jacobi constant is not exactly but approximately conserved\footnote{A similar result about the nonequivalence of PN Lagrangian and Hamiltonian approaches at the same order was discussed in \cite{WuPRD} and \cite{WuMNRAS}.} (see Section \ref{sec:3}).

For simplicity, in all that follows we shall use the canonical units introduced by \cite{Szebehely1967} for the normalization of constants in the CR3BP, that is
$$G m_1=1-\mu,\quad G m_2=\mu,\quad x_1=-\mu,\quad x_2=1-\mu,$$ 
where $\mu=m_{2}/(m_{1}+m_{2}) \in [0, 1/2]$. With this choice of units $a=1$, and the angular velocity \eqref{eq:omega} reduces to 
\begin{equation}\label{angm}
\omega=1+\frac{\mu(1-\mu)-3}{2 c^2}.
\end{equation}

\section{Region of validity of the PN approximations}\label{sec:2}

In this section, we propose a general criterion for the determination of approximate values of the mass-distance and velocity-speed of light ratios, which indicate the post-Newtonian approximation that should be appropriate according to the quotient.

For a binary system with masses $m_{1}$ and $m_{2}$, separated at a distance ${\cal R}$, with mean orbital velocity $v$, it can be shown that relative to $G(m_1+m_2)/(c^2 {\cal R})$ is the order of $(v/c)^2$ \citep{Blanchet2014}, 
\begin{equation}\label{eq:2.1}
\left(\frac{v}{c}\right)^{2 n} \propto \left(\frac{G {\cal M}}{c^2 {\cal R}}\right)^n ,
\end{equation}
where ${\cal M}=m_1+m_2$ is the total mass of the system and $n$ determines the post-Newtonian order to be considered. In the new system of units established by the Szebehely's definitions, the left-hand side of Eq. \eqref{eq:2.1} is related to the ratio ${\cal M/R}$ as
\begin{equation}\label{eq:2.2}
\left(\frac{v}{c}\right)^{2 n} \propto 1\times 10^{-8 n}\left(\frac{\cal M}{\cal R}\right)^n ,
\end{equation}
where ${\cal M}$ is measured in solar masses and ${\cal R}$ in astronomical units.

Now, let us impose the following criterion: The velocities ratio of the $n$-th PN-approximation becomes numerically negligible, i.e. of the order of the double machine precision $1\times10^{-16}$,  when this ratio is raised to the power of 2. Under this condition and from \eqref{eq:2.2} we get 
\begin{equation}\label{eq:2.3}
\frac{\cal M}{\cal R} \propto 1\times 10^{8 (n-1)/n},
\end{equation}
and hence the velocities ratio reads as
\begin{equation}\label{eq:2.4}
\frac{v}{c} \propto 1\times 10^{-4/n}.
\end{equation}

For the particular case $n=1$, considered in the present paper, and in accordance with \cite{Maindl1994}, from Eq. \eqref{eq:2.3} the quotient ${\cal M}$ divided by ${\cal R}$ is of the order 1 and from Eq. \eqref{eq:2.4} the quotient of the orbital velocity $v$ to the speed of light $c$ is of the order $10^{-4}$, then we may say that when considering general relativistic corrections, the $1$-PN approximation can appropriately be used to model binary systems composed by sun-planet couples of the solar system (see Table \ref{tab:2.1}) or binaries with masses around 1 solar mass orbiting at a distance of 1 AU. In table \ref{tab:2.2} we present the approximate range of validity of each PN-approximation in terms of the numerical values of ${\cal M/R}$ and $v/c$ for different values of $n$.
 
\begin{table}[t]
\caption{${\cal M/R}$, $v/c$ and $({\cal M/R})^{1/2} \div (v/c)$  ratios for each Sun-planet couple, with ${\cal M}$ expressed in solar masses and ${\cal R}$ in astronomical units.}
\begin{tabular}{@{}llll@{}}
\hline
System & ${\cal M/R}$ & $v/c$ & $({\cal M/R})^{1/2} \div (v/c)$\\
\hline
Mercury & 2.58 & 1.58$\times 10^{-4}$ & 1.02$\times 10^{4}$\\
Venus & 1.38 & 1.16$\times 10^{-4}$  & 1.01$\times 10^{4}$\\
Earth & 1.00 & 0.99$\times 10^{-4}$ & 1.01$\times 10^{4}$\\
Mars & 0.66 &0.80$\times 10^{-4}$ & 1.01$\times 10^{4}$\\
Jupiter & 0.19 & 0.44$\times 10^{-4}$ & 1.00$\times 10^{4}$\\
Saturn & 0.10 & 0.32$\times 10^{-4}$ & 0.99$\times 10^{4}$\\
Uranus & 0.05 & 0.23$\times 10^{-4}$ & 1.01$\times 10^{4}$\\
Neptune & 0.03 & 0.18$\times 10^{-4}$ & 1.01$\times 10^{4}$\\
\hline
\end{tabular}\label{tab:2.1}
\end{table}

\begin{table}[t]
\caption{${\cal M/R}$ and $v/c$ ratios calculated from Eqs. \eqref{eq:2.3} and \eqref{eq:2.4} respectively, for different values of $n$.}
\begin{tabular}{@{}llll@{}}
\hline
 PN-order & Approximation & ${\cal M/R}$ & $v/c$\\
\hline
$n=0$ &Newtonian & 0 & 0\\
$n=1$ &1-PN & 1 & $1\times 10^{-4}$\\
$n=2$ &2-PN & $1\times 10^{4}$ & $1\times 10^{-2}$\\
$n=3$ &3-PN & $1\times 10^{5}$ & $5\times 10^{-2}$\\
$n=4$ &4-PN & $1\times 10^{6}$ & $1\times 10^{-1}$\\
\vdots&\vdots&\vdots&\vdots\\
$n=\infty$&Gen. Rel. & $1\times 10^{8}$ & 1\\
\hline
\end{tabular}\label{tab:2.2}
\end{table}

It is important to note that the numerical values of the $v/c$ ratios presented in table \ref{tab:2.2}, calculated with the aid of Eq. \eqref{eq:2.4} are non-dimensional and hence independent of the system of units used.

As shown in table \ref{tab:2.2}, the main relativistic effects in the Solar system are $(v/c)^2$, which has an order of $10^{-8}$. Additionally, the column 4 in table \ref{tab:2.1} corresponds to the value of $c$, $10^{4}$. This value can also be derived in another path. $G^{*}{\cal M} =4\pi^{2}$ AU$^{3}$/yr$^2$ and $c^{*}= 299792458$ m/s =$6.31\times10^{4}$AU/yr (see also \citet{Klacka2008}). If ${\cal M}  =1$, then $G^{*}=4\pi^{2}$. We have $G^{*}/{c^{*}}^2 =0.9915\times10^{-8}\approx 10^{-8}$. However, the gravitational constant $G=1$ in the system of units established in this paper. In this case, the gravitational constant must be scaled as $G^{*}=4\pi^{2}\rightarrow G=1$. The speed of light should also be readjusted by $c^{*}=6.31\times10^{4}\rightarrow c$. That is, the term in the brackets in the right side of Eq. \eqref{eq:2.1} is readjusted as $4\pi^{2}/{c^{*}}^2\rightarrow G/c^2$. Thus, we have the readjusted speed of light $c=c^{*}/(2\pi) \approx 10^4$.

\section{Equivalence between the 1PN Lagrangian and Hamiltonian approaches}
\label{sec:4}

As noted in the introduction, the equivalence of the PN Lagrangian and Hamiltonian approaches is a topic of current debate. To provide further insight into this issue, here we shall compare the 1-PN Lagrangian equations of motion for the PCR3BP and the corresponding Hamiltonian equations of motion. To do so, let us start by considering the Lagrangian \eqref{Lf} in terms of the Newtonian ($\omega_{0}$) and Post-Newtonian ($\omega_{1}$) components of the angular velocity, which reads as

\begin{align}\label{eq:Lag5}
{\cal L}&=\frac{1}{2} \Omega_{0}^{2}+ G \left(\frac{m_{1}}{d_{1}} + \frac{m_{2}}{d_{2}}\right) +\frac{1}{c^2} \bigg\{ (x \dot{y}-\dot{x}y)\omega_{0} \omega_{1}  \nonumber\\ 
&-G^2\left[\frac{m_{1}m_{2}}{a}\left(\frac{1}{d_{1}} + \frac{1}{d_{2}}\right)
+\frac{1}{2}\left(\frac{m_{1}}{d_{1}} + \frac{m_{2}}{d_{2}}\right)^2\right]\nonumber\\ 
&+\frac{1}{8}\Omega_{0}^{4} +\frac{G}{2}\bigg[3\left(\frac{m_{1}}{d_{1}} + \frac{m_{2}}{d_{2}}\right)\Omega_{0}^{2}
- 7\omega_{0}(\dot{y}+\omega_{0} x) \nonumber\\ 
&\times\bigg(\frac{m_{1}x_{1}}{d_{1}}+\frac{m_{2}x_{2}}{d_{2}}\bigg)+ 3\omega_{0}^{2}\left(\frac{m_{1}x_{1}^{2}}{d_{1}} + \frac{m_{2}x_{2}^2}{d_{2}}\right)\nonumber\\ 
&- \omega_{0} y \left( \dot{x} - \omega_{0} y \right)\bigg[\frac{m_{1}x_{1}\left(x-x_{1}\right)}{d_{1}^{3}} + \frac{m_{2}x_{2}\left(x-x_{2}\right)}{d_{2}^{3}}\bigg] \nonumber\\
&- \omega_{0} y^{2} ( \dot{y} + \omega_{0} x)\left(\frac{m_{1}x_{1}}{d_{1}^{3}} + \frac{m_{2}x_{2}}{d_{2}^{3}}\right)\nonumber\\
&+ \omega_{0}^{2}  \omega_{1} (x^{2} + y^{2})
\bigg]
\bigg\},
\end{align}
 with $\Omega_{0}^{2}=\dot{x}^{2}+\dot{y}^2+\left(x^2+y^2\right)\omega_{0}^{2} +2 (x \dot{y}-\dot{x}y) \omega_{0}$. From \eqref{eq:pi} and the Lagrangian \eqref{eq:Lag5}, the components of the generalized momentum are given by 
\begin{align}\label{eq:px}
p_{x}&=\dot{x}-y\omega_{0}+ \frac{1}{c^2}\bigg\{\frac{1}{2} (\dot{x}-y\omega_{0})(\dot{x}^{2}+\dot{y}^2+\left(x^2+y^2\right)\omega_{0}^{2} \nonumber\\
&+2 (x \dot{y}-\dot{x}y) \omega_{0})+\frac{G}{2}\bigg[6\left(\frac{m_{1}}{d_{1}} + \frac{m_{2}}{d_{2}}\right)(\dot{x}-y\omega_{0})\nonumber\\
&-\bigg[\frac{m_{1}x_{1}\left(x-x_{1}\right)}{d_{1}^{3}} + \frac{m_{2}x_{2}\left(x-x_{2}\right)}{d_{2}^{3}}\bigg]y\omega_{0}\bigg]-y\omega_{0}\omega_{1}
\bigg\}
\end{align}
and
\begin{align}\label{eq:py}
p_{y}&=\dot{y}+x\omega_{0}+ \frac{1}{c^2}\bigg\{\frac{1}{2} (\dot{y}+x\omega_{0})(\dot{x}^{2}+\dot{y}^2+\left(x^2+y^2\right)\omega_{0}^{2} \nonumber\\
&+2 (x \dot{y}-\dot{x}y) \omega_{0})+\frac{G}{2}\bigg[6\left(\frac{m_{1}}{d_{1}} + \frac{m_{2}}{d_{2}}\right)(\dot{y}+x\omega_{0})\nonumber\\
&-7\bigg(\frac{m_{1}x_{1}}{d_{1}} + \frac{m_{2}x_{2}}{d_{2}}\bigg)\omega_{0}-\bigg(\frac{m_{1}x_{1}}{d_{1}^{3}} + \frac{m_{2}x_{2}}{d_{2}^{3}}\bigg)y^{2}\omega_{0}\bigg]\nonumber\\
&+x\omega_{0}\omega_{1}
\bigg\}.
\end{align}
Then, by solving equations \eqref{eq:px} and \eqref{eq:py} for $\dot{x}$ and $\dot{y}$,  and truncating higher-order PN terms, we get
\begin{align}\label{eq:xp}
\dot{x}&=p_{x}+y \omega_{0}+\frac{1}{c^2}\bigg\{y \omega_{0} \omega_{1}-3 G p_{x} \left(\frac{m_{1}}{d_{1}}+\frac{m_{2}}{d_{2}}\right)\nonumber\\
&+\frac{G}{2} \left[\frac{m_1 x_1 (x-x_1)}{d_{1}^3}+\frac{m_2 x_2 (x-x_2)}{d_{2}^3}\right]y \omega_{0} \nonumber\\
&-\frac{1}{2} p_{x} \left(p_{x}^2+p_{y}^2\right)   \bigg\},
\end{align}
and
\begin{align}\label{eq:yp}
\dot{y}&=p_{y}-x \omega_{0}+\frac{1}{c^2}\bigg\{-x \omega_{0} \omega_{1}-3 G p_{y} \left(\frac{m_{1}}{d_{1}}+\frac{m_{2}}{d_{2}}\right)\nonumber\\
&+\left[7\left(\frac{m_1 x_1}{d_{1}}+\frac{m_2 x_2}{d_{2}}\right)+
\left(\frac{m_1 x_1}{d_{1}^3}+\frac{m_2 x_2}{d_{2}^3}\right)y^2
\right]\nonumber\\
&\times\frac{G}{2} \omega_{0} -\frac{1}{2} p_{y} \left(p_{x}^2+p_{y}^2\right)   \bigg\}.
\end{align}

In section \ref{sec:1} we have shown that the Jacobi constant and the Hamiltonian satisfy the relation
$J = -{\cal H}$, thus from \eqref{Jacob} the Hamiltonian can be written as
\begin{align}\label{Hamiltonian}
{\cal H}&= \frac{1}{2}\left(p_{x}^{2}+p_{y}^{2}\right)+(y p_{x}-x p_{y})\omega_{0}-G \bigg(\frac{m_1}{d_1}+\frac{m_2}{d_2}\bigg)\nonumber\\
&+\frac{1}{c^2}\bigg\{\frac{G^2}{2}  \left[\left(\frac{m_{1}}{d_{1}}+\frac{m_{2}}{d_{2}}\right)^2 +\frac{2 m_1 m_2}{a} \left(\frac{1}{d_{1}}+\frac{1}{d_{2}}\right) \right]
\nonumber\\
&-\frac{3}{2} G (p_{x}^2+p_{y}^2)\left(\frac{m_{1}}{d_{1}}+\frac{m_{2}}{d_{2}}\right)
+(y p_{x}-x p_{y})\omega_{0}\omega_{1}
\nonumber\\
&
+\frac{G}{2} y p_{x} \omega_{0} \left[\frac{m_{1} x_{1}(x-x_{1})}{d_{1}^3}+\frac{m_{2} x_{2}(x-x_{2})}{d_{2}^3}\right]
\nonumber\\
&+\frac{7}{2} G p_{y} \omega_{0} \left(\frac{m_{1} x_{1}}{d_{1}}+\frac{m_{2} x_{2}}{d_{2}}\right)
-\frac{1}{8}(p_{x}^2+p_{y}^2)^2
\nonumber\\
&-\frac{3}{2} G \omega_{0}^2\left(\frac{m_{1} x_{1}^2}{d_{1}}+\frac{m_{2} x_{2}^2}{d_{2}}\right)
\nonumber\\
&+\frac{G}{2} p_{y} y^2 \omega_{0} \left(\frac{m_{1} x_{1}}{d_{1}^{3}}+\frac{m_{2} x_{2}}{d_{2}^{3}}\right)
\bigg\},
\end{align}
and its respective Hamilton's equations of motion are given by
\begin{align}
\dot{x}&=p_{x}+y \omega_{0}+\frac{1}{c^2}\bigg\{y \omega_{0} \omega_{1}-3 G p_{x} \left(\frac{m_{1}}{d_{1}}+\frac{m_{2}}{d_{2}}\right)\nonumber\\
&+\frac{G}{2} \left[\frac{m_1 x_1 (x-x_1)}{d_{1}^3}+\frac{m_2 x_2 (x-x_2)}{d_{2}^3}\right]y \omega_{0} \nonumber\\
&-\frac{1}{2} p_{x} \left(p_{x}^2+p_{y}^2\right)   \bigg\}, \label{eq:xpH}
\\
\dot{y}&=p_{y}-x \omega_{0}+\frac{1}{c^2}\bigg\{-x \omega_{0} \omega_{1}-3 G p_{y} \left(\frac{m_{1}}{d_{1}}+\frac{m_{2}}{d_{2}}\right)\nonumber\\
&+\left[7\left(\frac{m_1 x_1}{d_{1}}+\frac{m_2 x_2}{d_{2}}\right)+
\left(\frac{m_1 x_1}{d_{1}^3}+\frac{m_2 x_2}{d_{2}^3}\right)y^2
\right]\nonumber\\
&\times\frac{G}{2} \omega_{0} -\frac{1}{2} p_{y} \left(p_{x}^2+p_{y}^2\right),   \bigg\}. \label{eq:ypH}
\\
\dot{p}_x&= p_{y} \omega_{0} - G \left(\frac{ m_{1}
   (x- x_{1})}{d_{1}^{3}}+\frac{ m_{2} (x- x_{2})}{d_{2}^{3}}\right)\nonumber\\
&+\frac{1}{c^2} \bigg\{
\frac{G^2 m_{1}  m_{2}}{a} \left(\frac{x- x_1}{d_{1}^{3}}+\frac{x- x_2}{d_{2}^{3}}\right)\nonumber\\
&+G^2
   \left(\frac{ m_{1}^2 (x- x_{1})}{d_{1}^4}+ \frac{m_{2}^2 (x- x_{2})}{d_{2}^4}\right)\nonumber\\
&+\frac{G^2  m_{1} m_{2}}{d_{1} d_{2}} \left(\frac{x- x_{1}}{d_{1}^2}+\frac{x- x_{2}}{d_{2}^2}\right)   
\nonumber\\
&-\frac{3}{2} G \left(p_{x}^2+p_{y}^2\right) \left(\frac{m_{1} (x-x_{1})}{d_{1}^3}+\frac{m_{2} (x-x_{2})}{d_{2}^3}\right)
\nonumber\\
&+\frac{3}{2} G p_{x} y \omega_{0} \left(\frac{m_{1} x_{1} (x-x_{1})^2}{d_{1}^5}+\frac{m_{2} x_{2}
(x-x_{2})^2}{d_{2}^5}\right)
\nonumber\\
&+\frac{3}{2} G p_{y} y^2 \omega_{0} \left(\frac{m_{1} x_{1}
(x-x_{1})}{d_{1}^5}+\frac{m_{2} x_{2} (x-x_{2})}{d_{2}^5}\right)
\nonumber\\
&+\frac{7}{2} G p_{y} \omega_{0}
\left(\frac{m_{1} x_{1} (x-x_{1})}{d_{1}^3}+\frac{m_{2} x_{2} (x-x_{2})}{d_{2}^3}\right)
\nonumber\\
&-\frac{G}{2} p_{x} y \omega_{0} \left(\frac{m_{1} x_{1}}{d_{1}^3}+\frac{m_{2} x_{2}}{d_{2}^3}\right)+p_{y}\omega_{0}\omega_{1}
\nonumber\\
&-\frac{3}{2} G \omega_{0}^2 \left(\frac{m_{1} x_{1}^2 (x-x_{1})}{d_{1}^3}+\frac{m_{2} x_{2}^2
(x-x_{2})}{d_{2}^3}\right), \label{eq:pxpH}
\\
\dot{p}_y&= -p_{x} \omega_{0} - G y \left(\frac{ m_{1}}{d_{1}^{3}}+\frac{m_{2}}{d_{2}^{3}}\right)
\nonumber\\
&+\frac{1}{c^2} \bigg\{
G^2 y\left(\frac{m_{1}^{2}}{d_{1}^{4}}+\frac{m_{2}^2}{d_{2}^{4}}\right)
+\frac{G^2 m_{1} m_{2}}{a} y \left(\frac{1}{d_{1}^3}+\frac{1}{d_{2}^3}\right)
\nonumber\\
&-\frac{3}{2} G y \left(p_{x}^2+p_{y}^2\right)
\left(\frac{m_{1}}{d_{1}^3}+\frac{m_{2}}{d_{2}^3}\right)
+\frac{G^2 m_{1} m_{2}}{d_{1} d_{2}}  y 
\nonumber\\
&\times \left(\frac{1}{d_{1}^2}+\frac{1}{d_{2}^2}\right)+\frac{3}{2} G p_{y} y^3 \omega_{0} \left(\frac{m_{1} x_{1}}{d_{1}^5}+\frac{m_{2} x_{2}}{d_{2}^5}\right)
\nonumber\\
&+\frac{3}{2} G p_{x} y^2 \omega_{0}
\left(\frac{m_{1} x_{1} (x-x_{1})}{d_{1}^5}+\frac{m_{2} x_{2} (x-x_{2})}{d_{2}^5}\right)
\nonumber\\
&-\frac{1}{2} G p_{x} \omega_{0} \left(\frac{m_{1} x_{1} (x-x_{1})}{d_{1}^3}+\frac{m_{2} x_{2} (x-x_{2})}{d_{2}^3}\right)
\nonumber\\
&+\frac{5}{2} G p_{y} y \omega_{0} \left(\frac{m_{1} x_{1}}{d_{1}^3}+\frac{m_{2} x_{2}}{d_{2}^3}\right)
-p_{x}\omega_{0}\omega_{1}
\nonumber\\
&-\frac{3}{2} G y \omega_{0}^2 \left(\frac{m_{1} x_{1}^2}{d_{1}^3}+\frac{m_{2} x_{2}^2}{d_{2}^3}\right). \label{eq:pypH}
\end{align}

In order to show the equivalence between the Lagrangian and Hamiltonian equations of motion, we convert the first order system of equations (\ref{eq:xpH}-\ref{eq:pypH}) into a system of second order of the form $\ddot{\mathbf r}=f(\mathbf r,\dot{\mathbf r})$ by differentiating the velocities \eqref{eq:xpH} and \eqref{eq:ypH} with respect to time to find $\ddot{\mathbf r}$, and then substituting the derivatives of the momenta \eqref{eq:pxpH} and \eqref{eq:pypH}. It is worth mentioning that the resulting second-order differential equations are identical to the ones derived by \cite{Maindl1994}, for this reason, in all that follows we will refer indistinctly to the Hamiltonian system and to the Maindl's equations of motion.

\begin{figure}[h!]
\centering
\begin{tabular}{cc}
\includegraphics[width=0.4\textwidth]{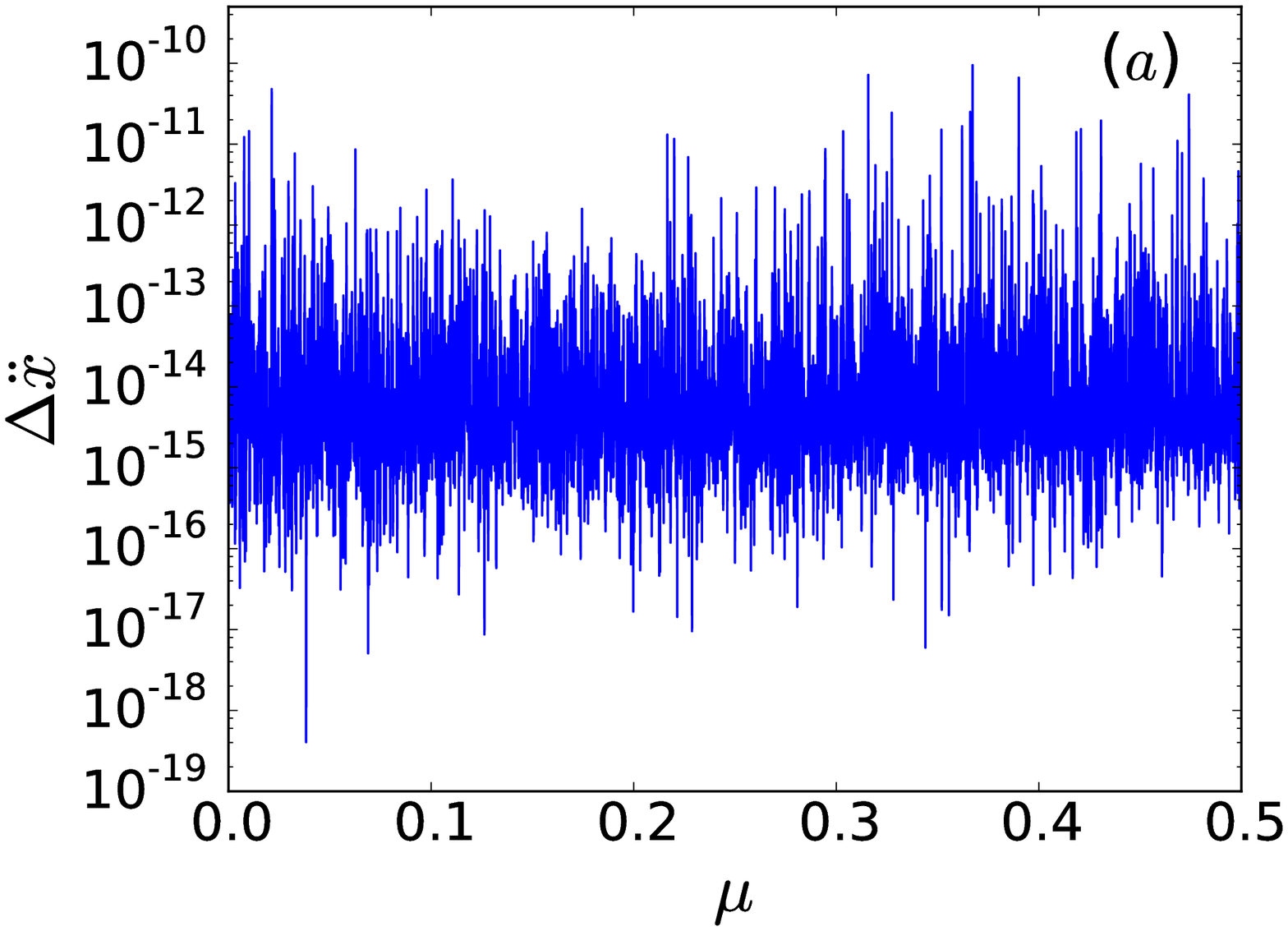}\\
\includegraphics[width=0.4\textwidth]{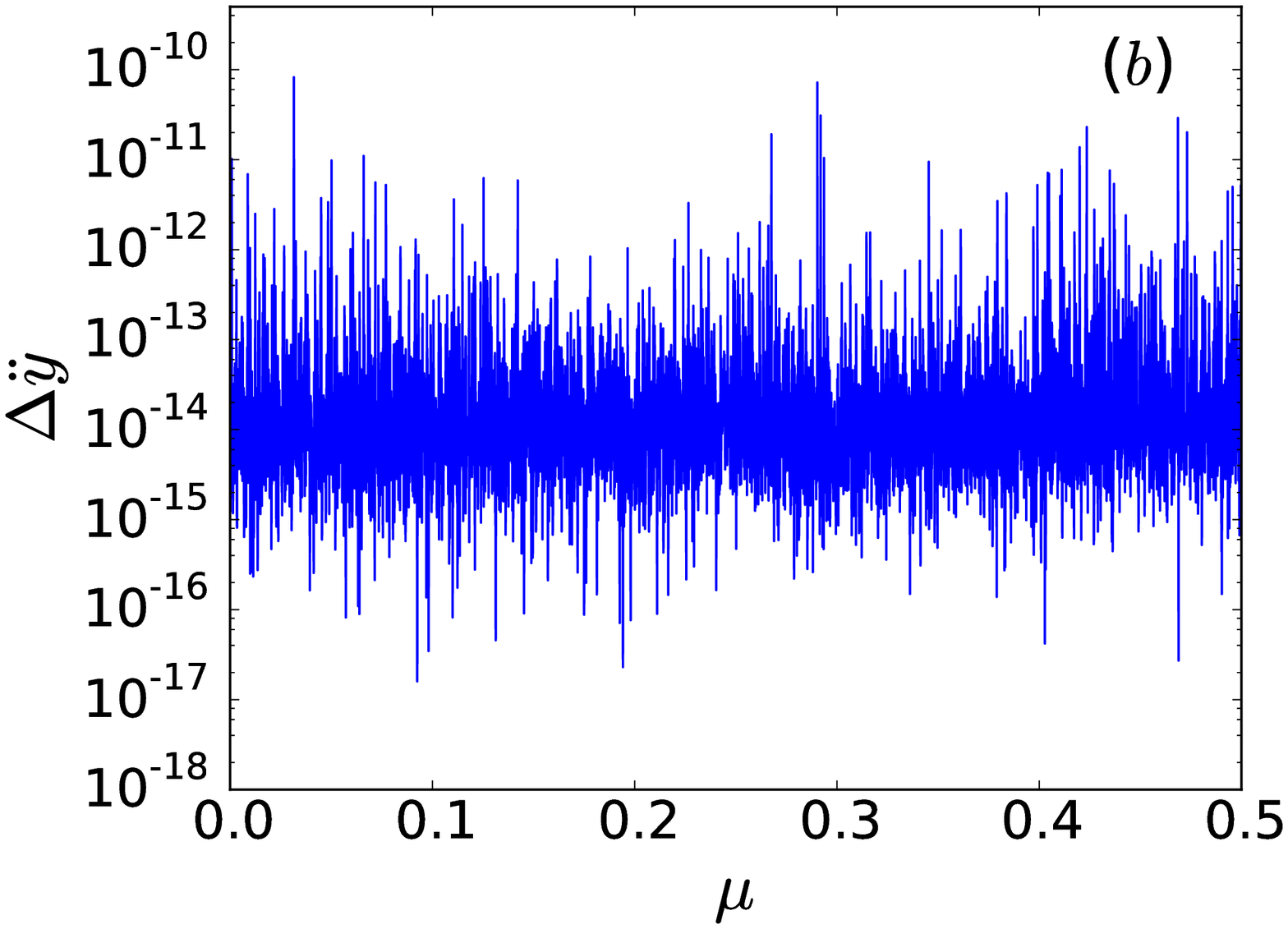}
\end{tabular}
\caption{Equivalence between the Lagrangian and Hamiltonian representations in the 1-PN PCR3BP. (a) Differences for the equations of motion in $x$, $\Delta \ddot{x}= \ddot{x}_{L}- \ddot{x}_{H}$, (b) Differences for the equations of motion in $y$, $\Delta \ddot{y}= \ddot{y}_{L}- \ddot{y}_{H}$. The subindex denotes the used formalism, {\it i.e} L for Lagrangian and H for Hamiltonian. The parameters $x,y,\dot{x}$, and $\dot{y}$, were generated randomly for each $\mu$.}
\label{fig6}
\end{figure}

In figure \ref{fig6}, we present the numerical values of the differences between the equations of motion derived from the Lagrangian and Hamiltonian formalisms, in terms of the mass parameter $\mu$. The numerical values of the positions and velocities $(x,y,\dot{x},\dot{y})$ were obtained by means of a random number generator, with $c=10^{4}$ according to the PN approximation and the unit system being used. It should be pointed out that only in the Newtonian limit $c\rightarrow\infty$, the Lagrangian and Hamiltonian equations of motion are identical.

The observed differences have a mean value of the order $10^{-15}$, {\it i.e} proportional to the next to leading post-Newtonian order $\propto 1/c^{4}$, which could indicate that the claimed differences between both approaches are due to an inappropriate setting of the speed of light in a given system of units. In fact, if we set $c=1$ the resulting differences are of the order $10^{1}$. In other words, with a right setting of parameters, the Lagrangian and Hamiltonian equations of motion are not identical but are approximately equivalent. 

\section{Results}
\label{sec:3}

By taking the total time derivative of the Jacobi constant Eq. (\ref{Jacob}), 
\begin{equation}\label{eq:dJ}
\frac{dJ}{dt}=\dot{x}\frac{\partial J}{\partial x}+\dot{y}\frac{\partial J}{\partial y}+\ddot{x}\frac{\partial J}{\partial \dot{x}}+\ddot{y}\frac{\partial J}{\partial \dot{y}},
\end{equation}
and substituting Eq. \eqref{eq:xppf} and Eq. \eqref{eq:yppf} into Eq. \eqref{eq:dJ}, it can be shown that  
\begin{equation}\label{eq:dif}
\frac{dJ}{dt}\approx c^{-4} h(x,y,\dot{x},\dot{y};\mu),
\end{equation}
then, the accuracy of the approximations is determined by an appropriate choice of the constant $c$ and can be verified through inspection of the conservation of the Jacobi constant. 

\begin{figure}[h!]
\centering
\begin{tabular}{c}
\includegraphics[width=0.4\textwidth]{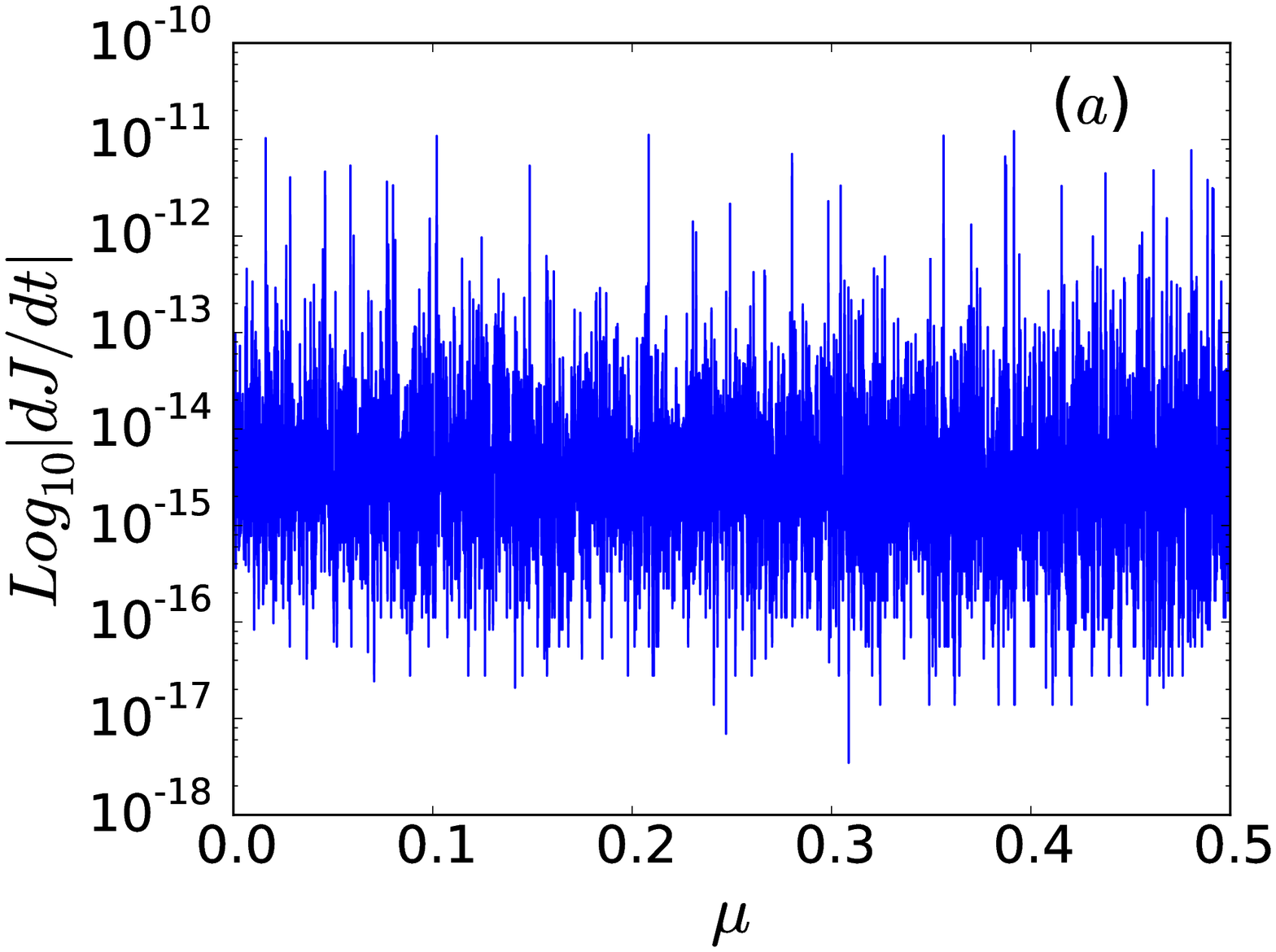}\\
\includegraphics[width=0.4\textwidth]{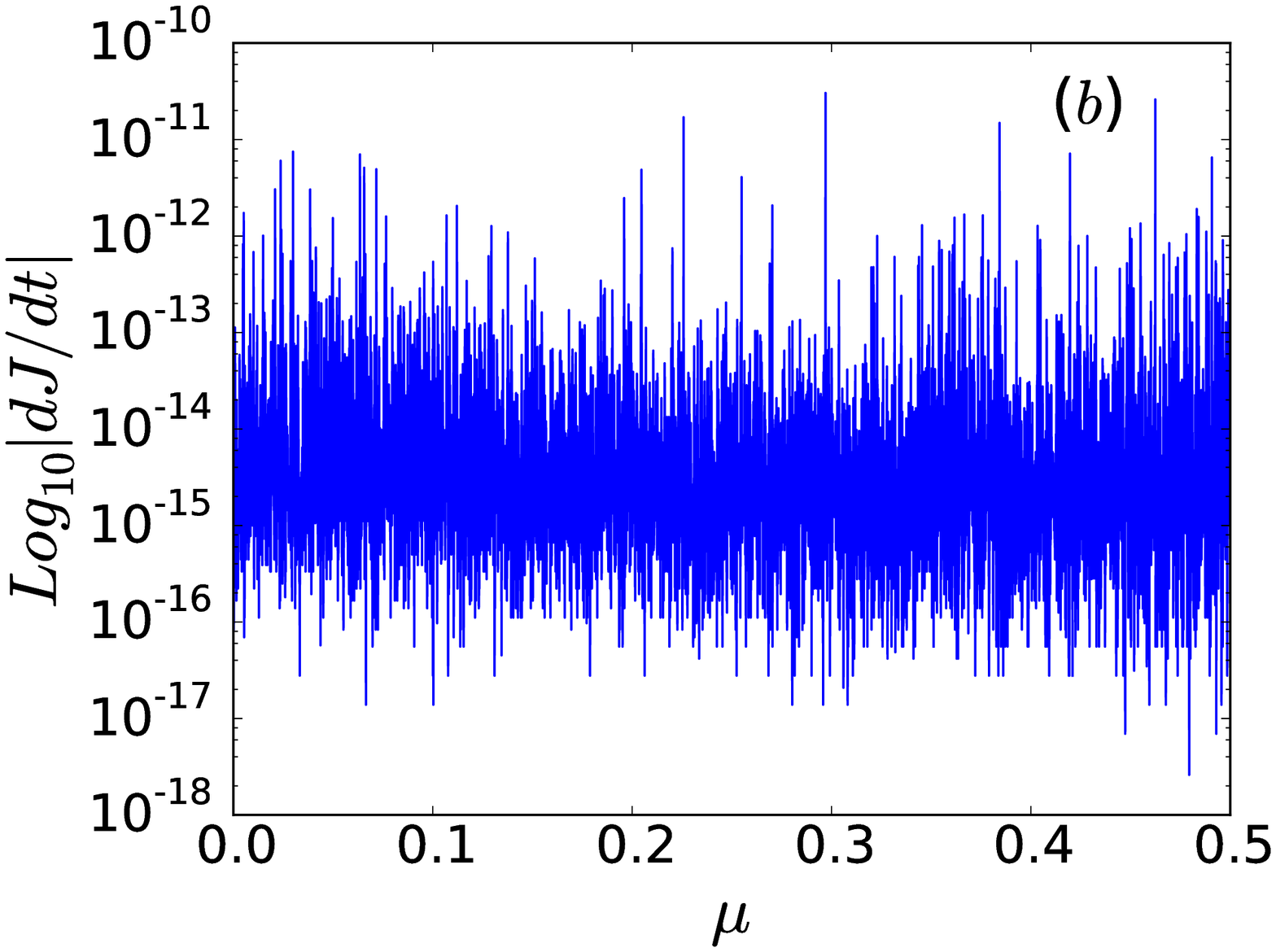}\\
\includegraphics[width=0.4\textwidth]{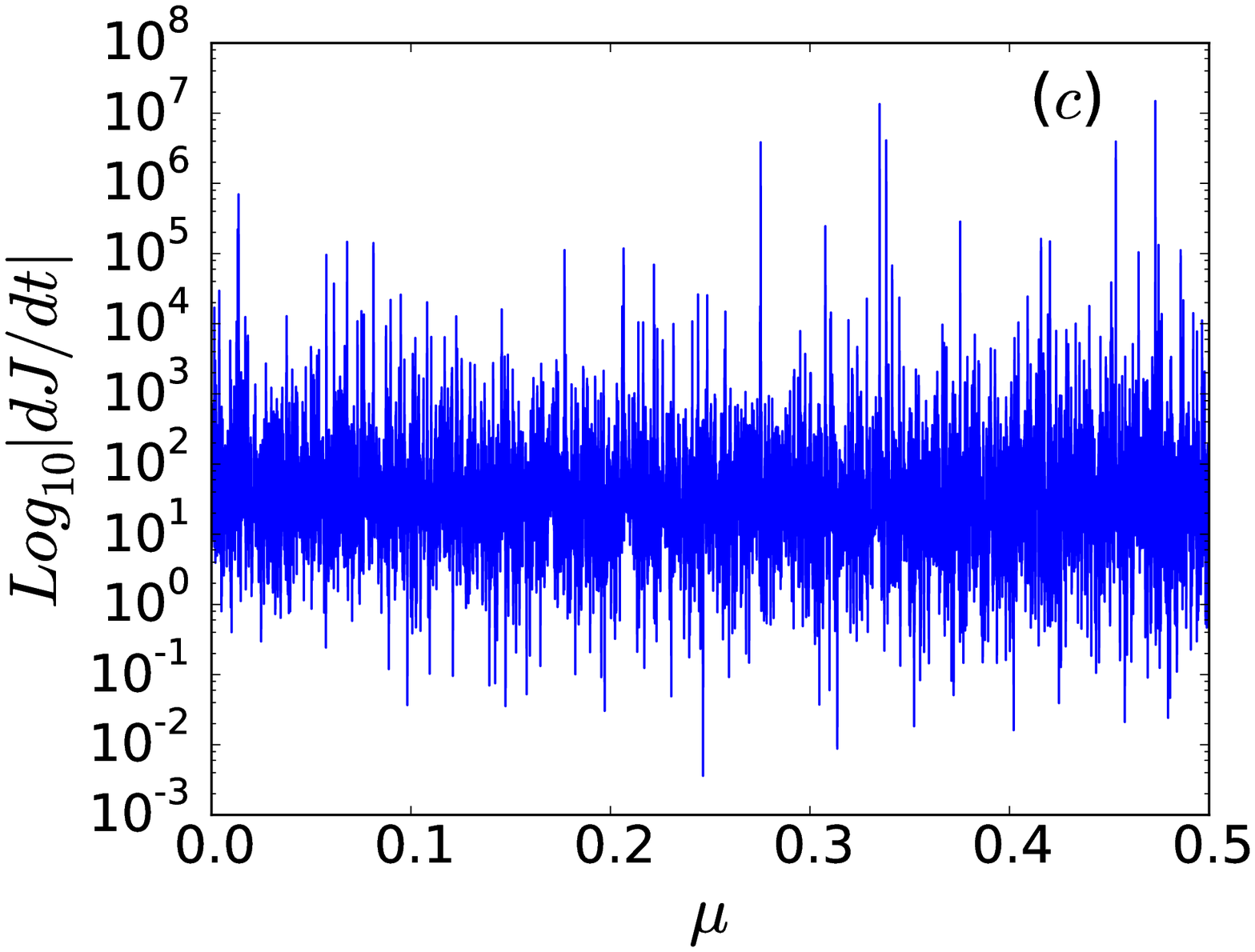}
\end{tabular}
\caption{Conservation of the Jacobi constant in the 1-PN PCR3BP using the expressions (a) derived derived from the Hamiltonian formalism, (b) the ones derived from the Lagrangian formalism and (c) the setup used by \cite{Huang2014}. The parameters $x,y,\dot{x}$, and $\dot{y}$, were generated randomly for each $\mu$.}
\label{fig1}
\end{figure}

As far as we know, the first and only study on the dynamics of the 1-PN PCR3BP was performed by Huang and Wu in \citeyear{Huang2014}. However, in their work the authors set $c=1$, which would correspond to a rather extreme system outside the scope of the 1-PN approximation. In order to clarify the last point, in figure \ref{fig1} we present the numerical values of the total time derivative of the Jacobi constant Eq. \eqref{eq:dJ} in terms of the mass parameter $\mu$. The values of $dJ/dt$ using the Jacobi constant together with the Hamiltonian equations of motion \citep{Maindl1994} are presented in figure \ref{fig1} (a), while the results for the Jacobi constant and the Lagrangian equations of motion (\ref{eq:xppf}-\ref{eq:yppf}) are shown in figure \ref{fig1} (b). In this two cases, the constant $c$ was taken as $c=1\times 10^{4}$. Moreover, in figure \ref{fig1} (c), the values of $dJ/dt$ are presented for the setup used by \cite{Huang2014} with $c=a=1$. In the three cases the numerical values of the positions and velocities $(x,y,\dot{x},\dot{y})$ were obtained by means of a random number generator.

The results presented in Fig. \ref{fig1} suggest that the choice $c = 1$ is inadequate since the Jacobi constant of the 1-PN approximation is not conserved, leading to spurious results on the dynamics of the system.\footnote{For the case $c=1$ in \cite{Huang2014}, the Jacobi constant still has higher accuracy when $a=1$ and the velocity at any time is smaller than $10^{-4}$, or when the distance $a$ takes a larger value (see \cite{Su2016} and \cite{Chen2016}).}  Also, it is important to note that the results obtained for Fig. \ref{fig1} (a) and Fig. \ref{fig1} (b) are very similar, which is obvious if taking into account that the equations of motion are not the same, but are equivalent. 

Once we have proved that the Jacobi constant is preserved to a good approximation in our approach, it is possible to analyze the dynamics of the post-Newtonian system by means of the Poincar\'e section method, {\it i.e} the constant of motion $J$ allows us to effectively reduce the $4$-dimensional phase space into a $3$-dimensional one, hence the Poincar\'e section, which is one dimension smaller than the considered system, permit us to analyze in a straightforward manner the dynamics of the original system in a $2$-dimensional phase plane.

\begin{figure}[h!]
\centering
\begin{tabular}{cc}
\includegraphics[width=0.40\textwidth]{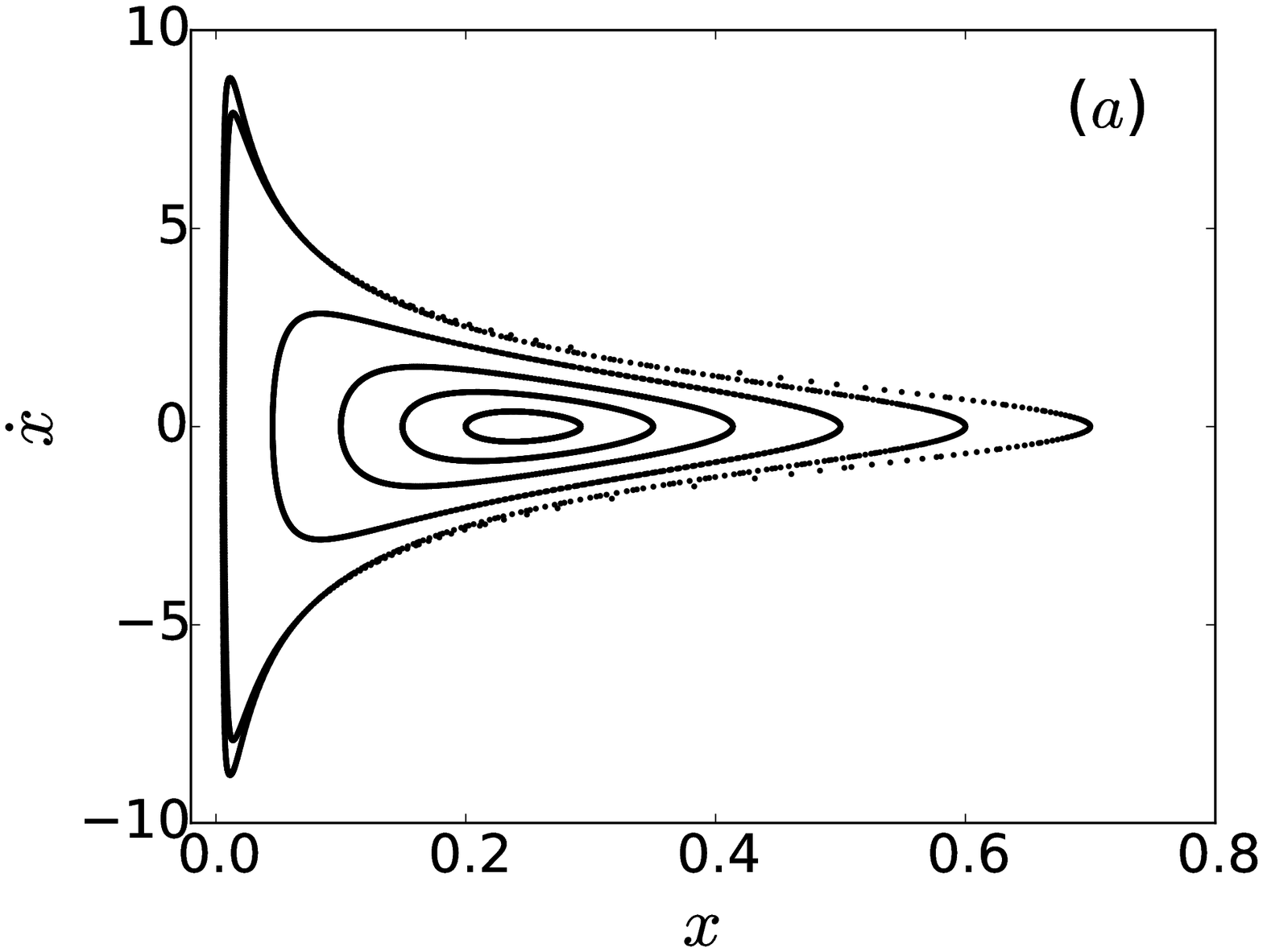}\\
\includegraphics[width=0.40\textwidth]{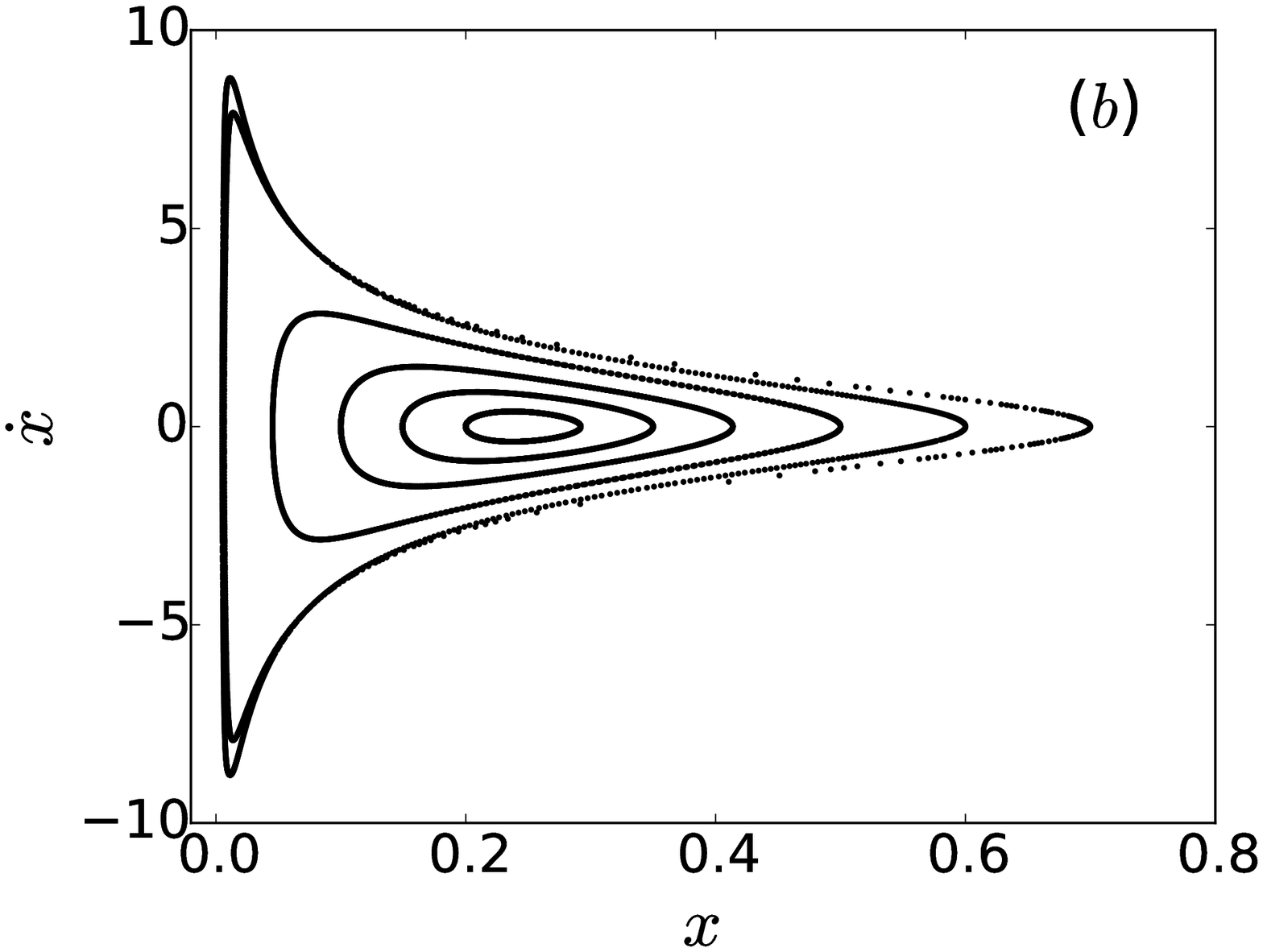}
\end{tabular}
\caption{Typical Poincar\'e surface of section for the system of Eqs. \eqref{eq:xppf} and \eqref{eq:yppf}. We used $c\rightarrow\infty$ for the classical system (a) and $c=10^{4}$ for the post-Newtonian one (b). The other parameters have been set to $\mu=10^{-3}$ and $J=1.535$. Areas of regular motion with periodic and quasi-periodic orbits are present in both regimes.}
\label{fig2}
\end{figure}

\begin{figure}[h!]
\centering
\begin{tabular}{cc}
\includegraphics[width=0.40\textwidth]{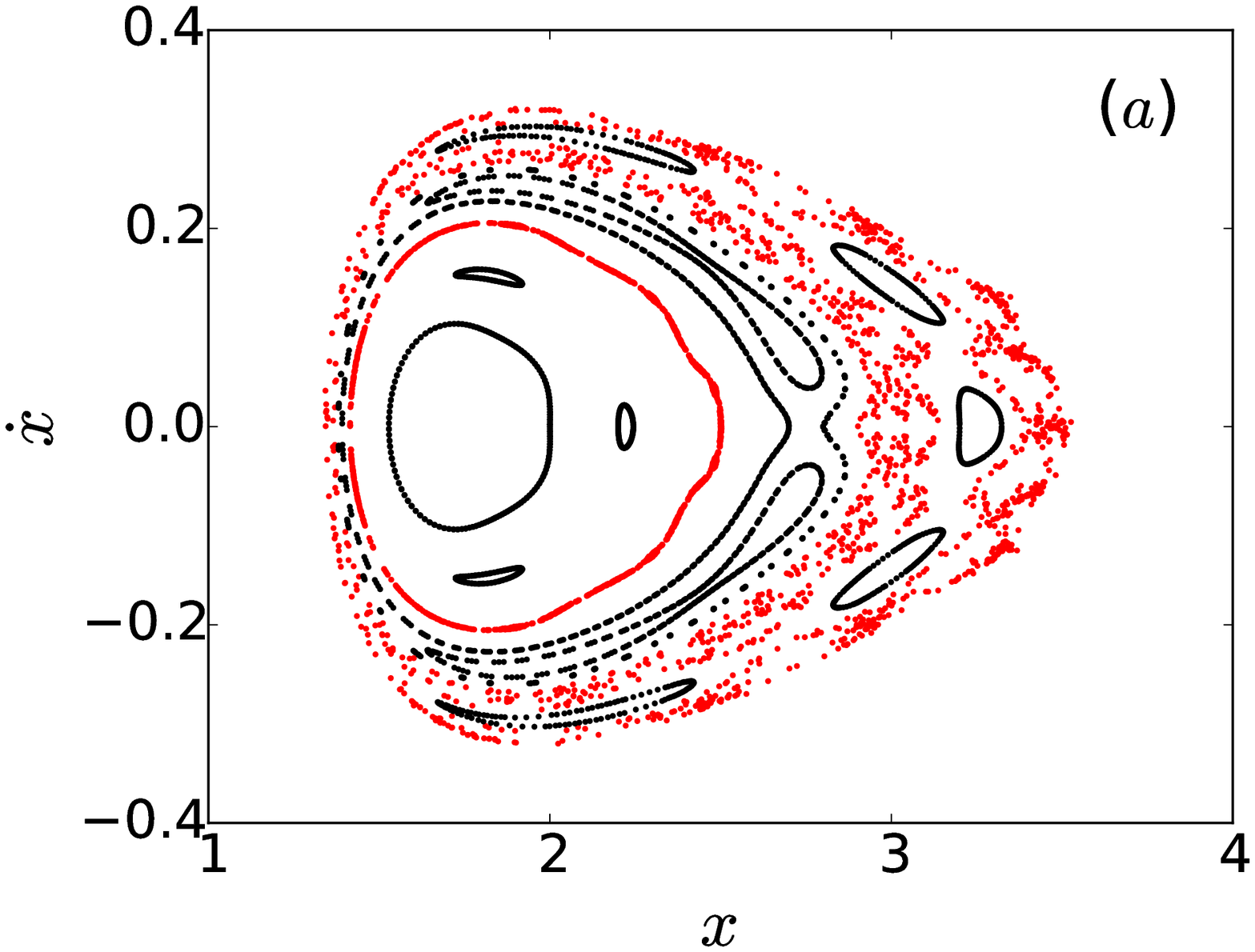}\\
\includegraphics[width=0.40\textwidth]{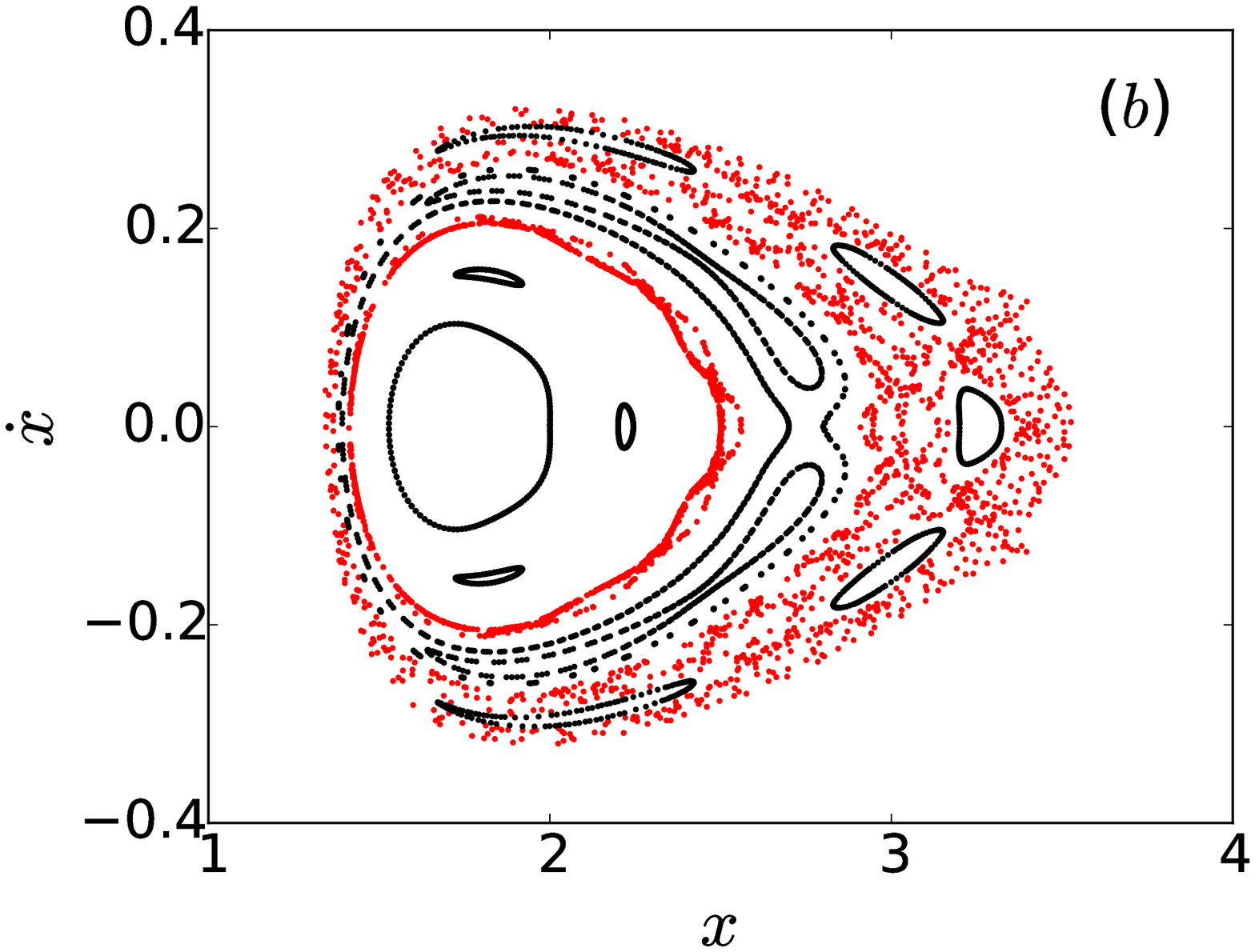}
\end{tabular}
\caption{Typical Poincar\'e surface of section for the system of Eqs. \eqref{eq:xppf} and \eqref{eq:yppf}. We used $c\rightarrow\infty$ for the classical system (a) and $c=10^{4}$ for the post-Newtonian one (b). The other parameters have been set to $\mu=10^{-3}$ and $J=1.6$. Areas of regular motion (black) coexisting with chaotic (red) ones are present in both regimes.}
\label{fig3}
\end{figure}

In Figs. \ref{fig2} and \ref{fig3}, we show two typical Poincar\'e surfaces of section for the Newtonian system (Panels a) and their post-Newtonian counterpart (Panels b), using two different values of the Jacobi constant. The equations of motion have been integrated using a Runge-Kutta-Fehlberg 7(8) method of integration with automatic step-size control. With this method, the relative error in the integral of motion is lower than $10^{-9}$ over a time span of $10^4$ in the integration time. In all the cases, we choose the condition $y=0$ with $\dot{y}<0$ to represent the surface of sections into the $(x,\dot{x})$-plane.

We show in Fig. \ref{fig2} an example of a Poincar\'e section using a mass parameter $\mu=10^{-3}$ and Jacobi constant $J=1.535$, for six different initial conditions. It can be observed that the results for the Newtonian (a) and post-Newtonian (b) surfaces of section are practically identical when the phase space is filled with definite periodic and quasi-periodic orbits. On the other hand, in Fig. \ref{fig3} we used $\mu=10^{-3}$ and $J=1.6$ for nine different initial conditions, the chaotic orbits are plotted in red color while the regular ones are plotted in black color. As can be noted, the transition from the classical (a) to the post-Newtonian (b) regimes, exhibits non-negligible variations only for the chaotic orbits. This result was verified using different values of the Jacobi constant, observing the same tendency.

\begin{figure}[h!]
\centering
\begin{tabular}{cc}
\includegraphics[width=0.4\textwidth]{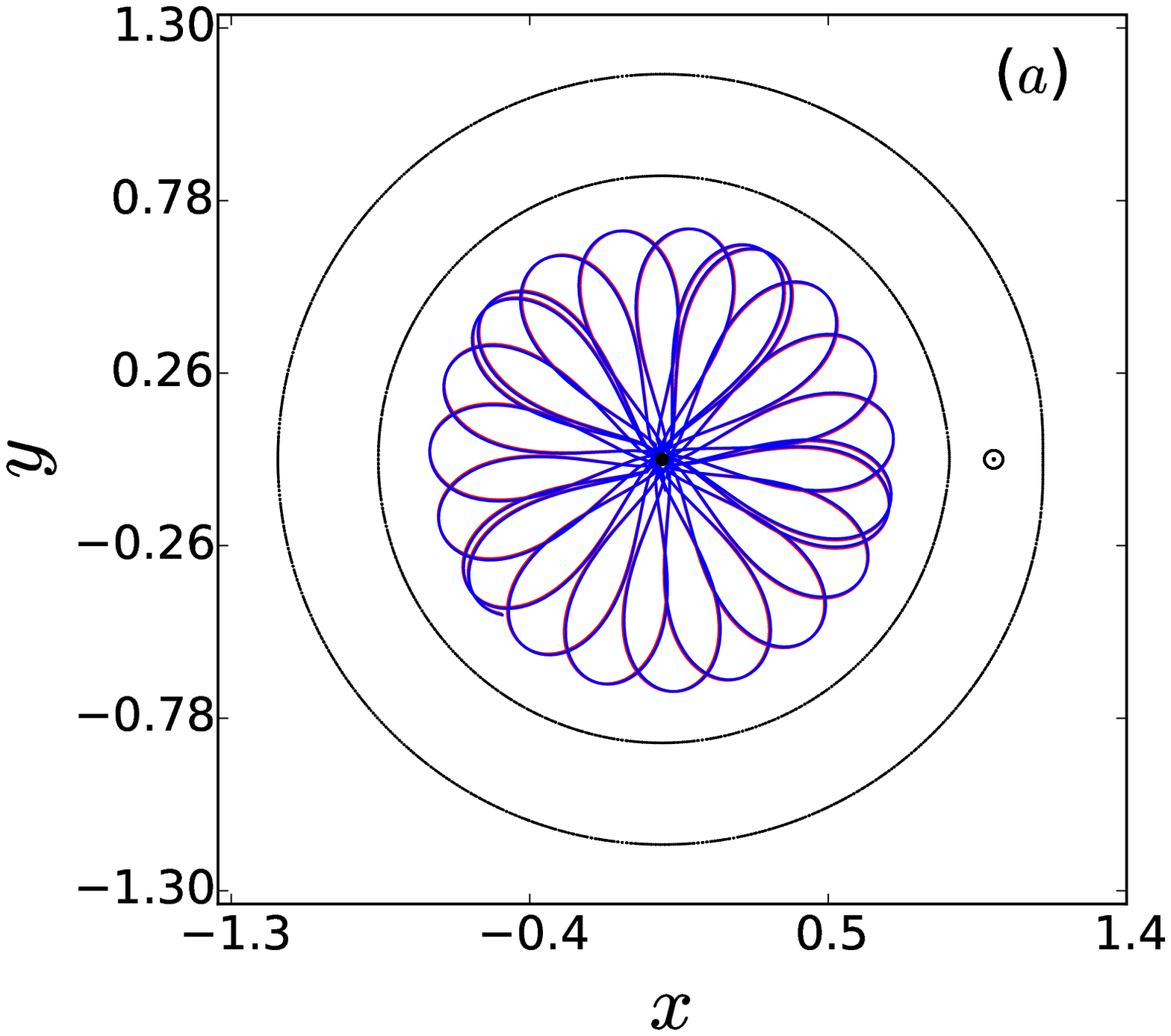}\\
\includegraphics[width=0.4\textwidth]{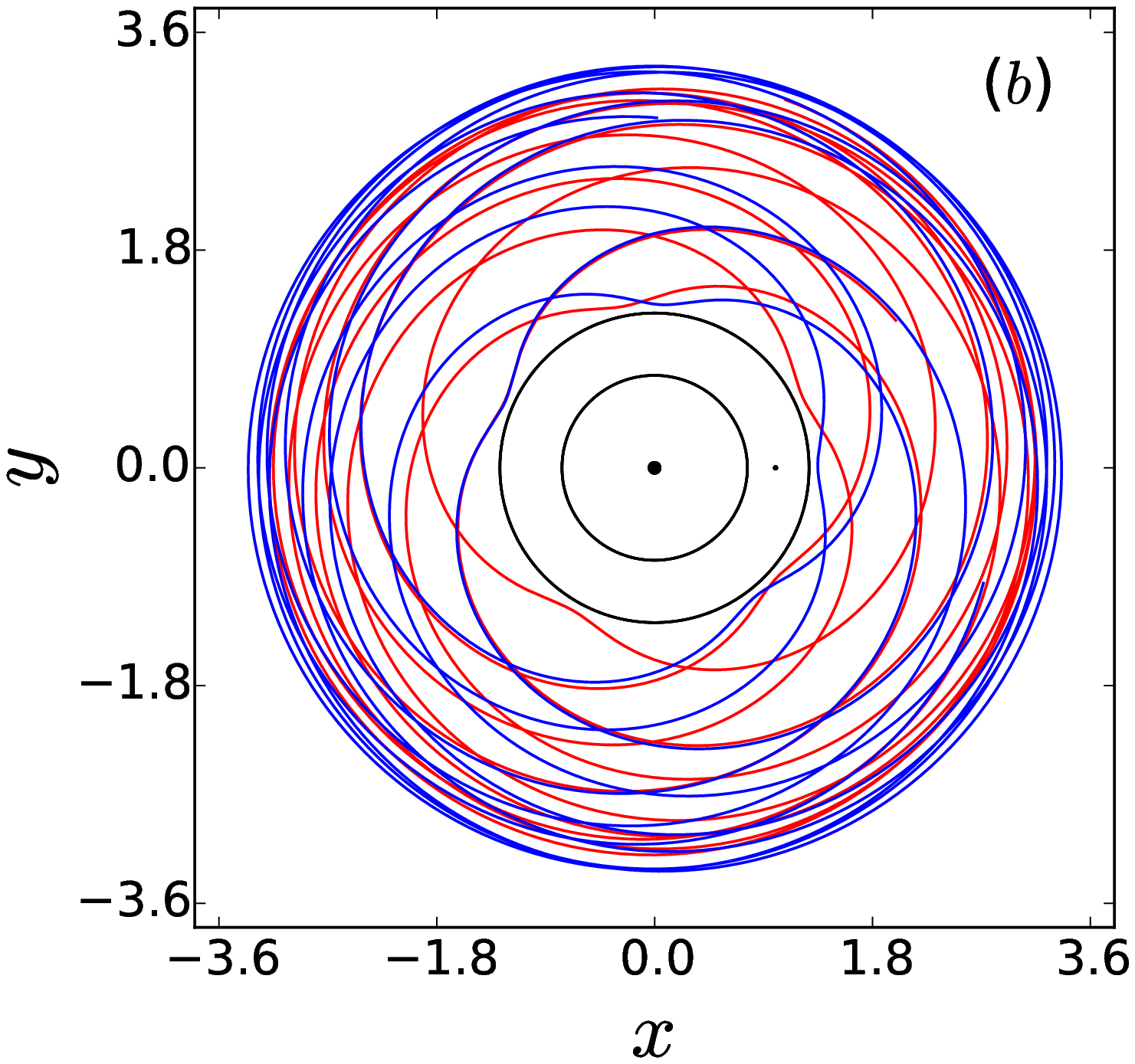}
\end{tabular}
\caption{Evolution of classical trajectories (red curves) and their corresponding post-Newtonian counterparts (blue curves). The panel (a) corresponds to the superposition of regular orbits with initial conditions $x_{0}=0.7, y_{0}=0, \dot{x_{0}}=10^{-4}$ and Jacobi constant $J=1.535$. The panel (b) corresponds to the superposition of chaotic orbits with initial conditions $x_{0}=2.95, y_{0}=0, \dot{x_{0}}=10^{-4}$ and Jacobi constant $J=1.6$. The central dots denote the position of primaries whose radius is proportional to its mass.  The black contours denotes the Hill region.}
\label{fig4}
\end{figure}

The main feature of chaotic orbits is the divergence of nearby trajectories, which is reflected by the strong dependence on initial conditions. In Fig. \ref{fig4} (a), we consider the evolution of two regular trajectories for the system (\ref{eq:xppf}-\ref{eq:yppf}) using $c\rightarrow\infty$ for the classical Newtonian system (in red color) and  $c=10^{4}$ for the post-Newtonian one (in blue color), with initial conditions $x_{0}=0.7, y_{0}=0, \dot{x_{0}}=10^{-4}$ and Jacobi constant $J=1.535$. On the other hand, in figure \ref{fig4} (b) we consider two chaotic orbits using the same color-coding, with initial conditions $x_{0}=2.95, y_{0}=0, \dot{x_{0}}=10^{-4}$ and Jacobi constant $J=1.6$. In each case, the numerical value of $\dot{y_{0}}$ is determined from Eq.  \eqref{Jacob}. In the case of regular dynamics, after a long time evolution, the classical and post-Newtonian orbits are practically indistinguishable (see Fig. \ref{fig4} (a)), but for the chaotic motion, the sensitivity to initial conditions leads to a substantial deviation in the trajectories (see Fig. \ref{fig4} (b)). In order to clearly appreciate the separation among orbits, we have plotted the orbits for the interval of time $t \in [6900,7000]$. 

From the spread of the chaotic zone observed in Fig. \ref{fig3}, we may infer that there exist an increase in the chaoticity of the post-Newtonian system compared to its Newtonian counterpart. This hypothesis will be proved with the aid of the largest Lyapunov exponent $\lambda_{max}$, which measures the mean exponential rate of divergence of nearby trajectories. Two representative approaches for the calculation of $\lambda_{max}$ are the variational method and the two-particle method, both methods  were compared by \cite{Tancredi2001}, showing that the two-particle method can lead to a false estimation of a positive $\lambda_{max}$. In the context of general relativity, the invariance of the Lyapunov exponents has been an intense area of research (see {\it e.g.} \cite{Motter2009}) and some gauge invariant indicators have been introduced (see \cite{Wu2003}; \cite{Wu2006}; \cite{Lukes2014}). In what follows we shall use the variational method because it has been previously shown that this approach is very accurate for a wide class of dynamical systems \citep{Dubeibe2014}. In this case, a bounded orbit is said to be chaotic for $\lambda_{max}>0$, or regular for $\lambda_{max}=0$. 

\begin{figure}[h!]
\centering
\begin{tabular}{cc}
\includegraphics[width=0.4\textwidth]{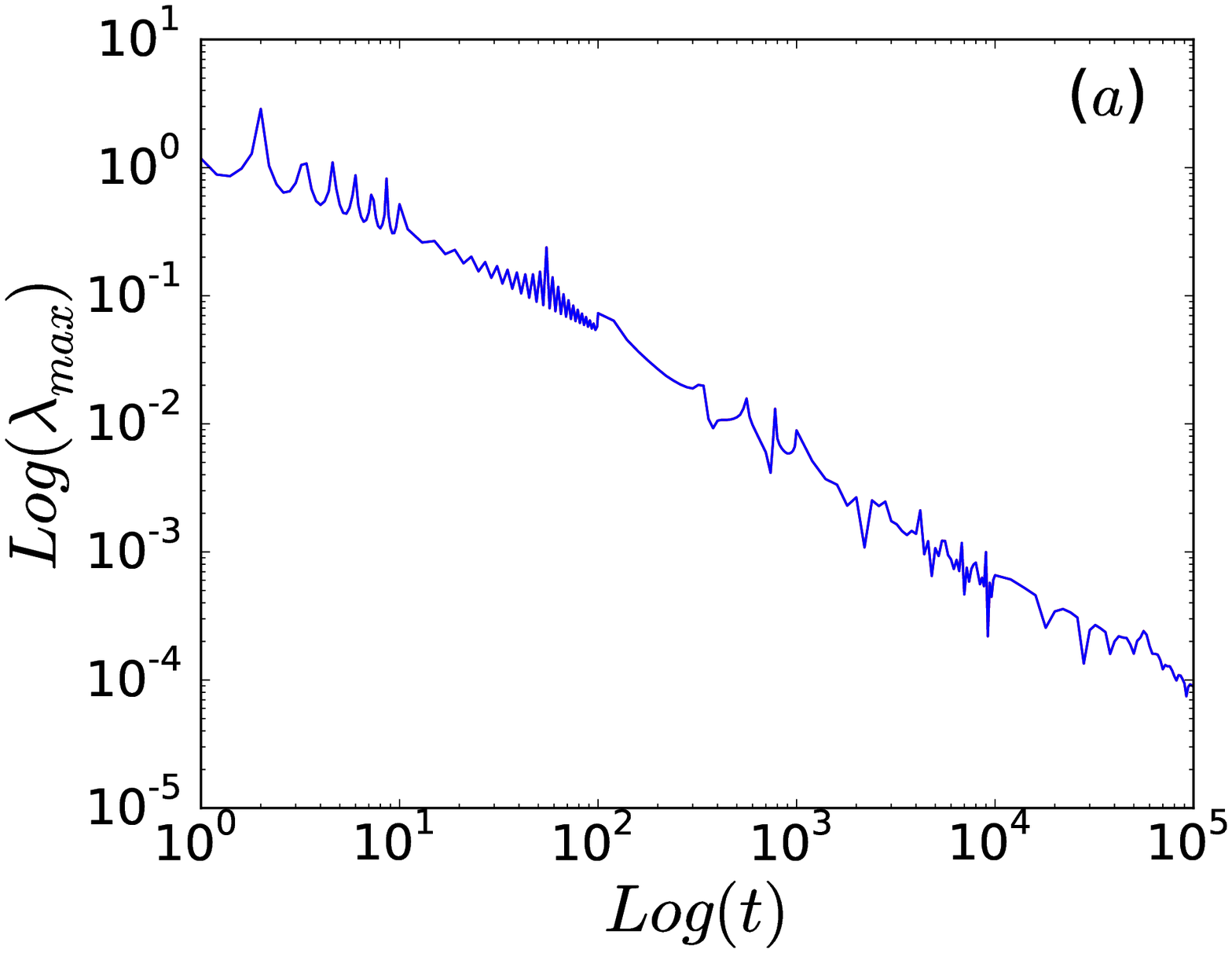}\\
\includegraphics[width=0.4\textwidth]{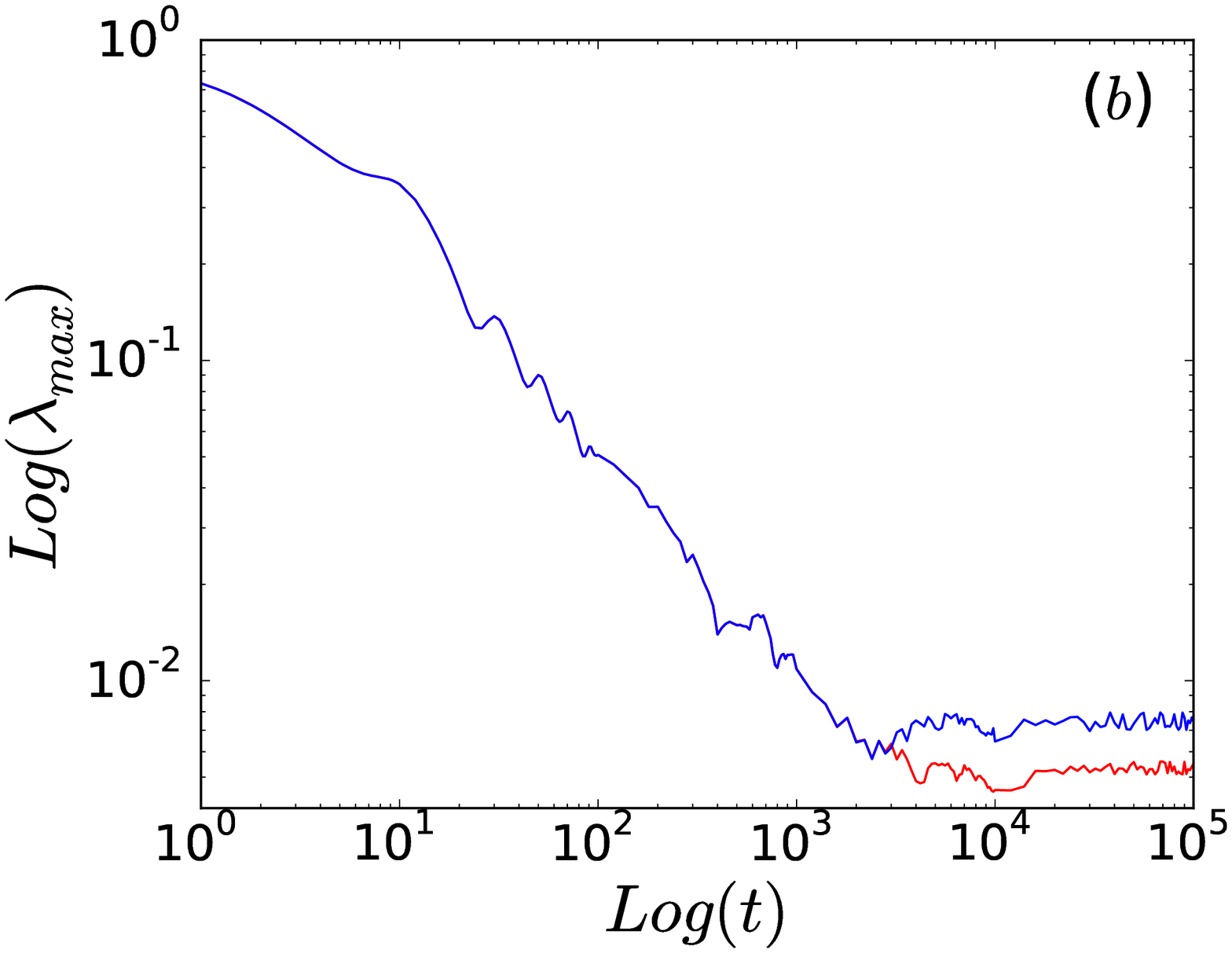}
\end{tabular}
\caption{Largest Lyapunov exponent for the system (\ref{eq:xppf}-\ref{eq:yppf}) with $c\rightarrow\infty$ (red curves) and  $c=10^{4}$ (blue curves). The panel (a) corresponds to the superposition of $\lambda_{max}$ for regular orbits with initial conditions $x_{0}=0.7, y_{0}=0, \dot{x_{0}}=10^{-4}$ and Jacobi constant $J=1.535$. The panel (b) corresponds to the superposition of $\lambda_{max}$ for chaotic orbits with initial conditions $x_{0}=2.95, y_{0}=0, \dot{x_{0}}=10^{-4}$ and Jacobi constant $J=1.6$.}
\label{fig5}
\end{figure}

In Fig. \ref{fig5}, we show a comparison of $\lambda_{max}$ in the Newtonian (red curve) and post-Newtonian (blue curve) regimes, for the corresponding initial conditions used in Fig. \ref{fig4}. From Fig. \ref{fig5} (a) the rate of divergence of nearby trajectories shows a polynomial distribution, as is expected for a regular system.  
Moreover, the logarithmic plots of the average largest Lyapunov exponent for the Newtonian and the 1-PN regimes, are exactly alike. Finally in Fig. \ref{fig5} (b), after a sufficiently long integration time, the values of $\lambda_{max}$ tend to a value greater than zero, which undoubtedly indicate the chaoticity of the orbits in the Newtonian (red curve) and the 1-PN (blue curve) regimes. As can be noted, the chaotic indicator is slightly increased for the post-Newtonian system in comparison to the Newtonian one.  

\section{Conclusions}
\label{sec:5}

In the present paper, using the Lagrangian formalism, we have derived an alternative new system of equations of motion for the planar circular restricted three-body problem at the first post-Newtonian approximation order. The corresponding Hamilton's equations of motion were shown to be equivalent but non-identical to the former system, mainly due to the truncation of higher-order PN terms. We also have introduced two formulas, showing that in a binary system, the quotient of the mean orbital velocity divided by the speed of light, must take values of the order $10^{-4}$, in order to be consistent with the first-order post-Newtonian approximation. As a natural result emerges the possibility to study the dynamics of the first-order post-Newtonian system in terms of the separation of the primaries, as long as the value of the total mass-to-separation ratio is approximately one (with the mass and distance measured in Solar masses and Astronomical units, respectively). Moreover, if the last conditions are fulfilled, it can be concluded from the numerical simulations that the successive approximations performed to derive the equations of motion, turn the Jacobi constant not exactly, but approximately conserved.

On the other hand, by using the Poincar\'e section method and the largest Lyapunov exponent, we have also studied the dynamics of the equations of motion, where it was found that the system can be either regular or chaotic. For specific values of the Jacobi constant most of the orbits exhibit regular dynamics, while in the other cases the dynamic is mainly mixed, {\it i.e.} regular orbits embedded in a chaotic sea. None of the studied cases showed a fully chaotic phase space. 

In general, our numerical results suggest that the first-order post-Newtonian approximation may be understood as a small perturbation of the Newtonian system, in the sense that the post-Newtonian corrections are of the order $10^{-8}$. Due to the properties of the nonlinear dynamical systems, the Newtonian and post-Newtonian orbits are appreciably different only in the chaotic regime. As a novel property, we observed a chaotic amplification effect, {\it i.e.} the post-Newtonian orbits exhibit an increased chaoticity with respect to the Newtonian ones. Our results could have significant implications for the study of the dynamics of the Solar System, since the characteristic timescales (Lyapunov times) are expected to be reduced when considering the first-order post-Newtonian corrections.

Finally, it is important to note that there is no explicit conflict between the article of \cite{Huang2014} and the present paper, in fact, the results of the present paper are complementary to those obtained by \cite{Huang2014}. This is because the scaling transformation together with the choice of $c=1$ in \cite{Huang2014} is suitable for the case of $a\gg 1$, while $c=10^4$ is suitable for the case of $a=1$ in the post-Newtonian circular restricted three-body problem.

\acknowledgments
We thank the anonymous referee for constructive criticisms and suggestions that helped us improve this paper.
FLD acknowledges financial support from Universidad de los Llanos, under Grants Commission: Postdoctoral Fellowship Scheme. FDLC and GAG gratefully acknowledges the financial support provided by VIE-UIS, under grants numbers 1822, 1785 and 1838, and COLCIENCIAS, Colombia, under Grant No. 8840.

\bibliographystyle{spr-mp-nameyear-cnd}

\begin{thebibliography}{99}

\bibitem[Blanchet(2014)]{Blanchet2014} Blanchet, L.: Gravitational radiation from post-Newtonian sources and inspiralling compact binaries. Living Reviews in Relativity {\bf 17} 2 (2014).

\bibitem[Bombardelli \& Pel\'aez(2011)]{Bombardelli2011} Bombardelli, C. and Pel\'aez J.: On the stability of artificial equilibrium points in the circular restricted three-body problem. Celestial Mechanics and Dynamical Astronomy {\bf 109} (1), 13 (2011).

\bibitem[Brumberg(1972)]{Brumberg1972} Brumberg, V.A. Relativistic Celestial Mechanics. Moscow (1972); Brumberg, V.A. Essential Relativistic Celestial Mechanics. Hilger, Bristol (1991).

\bibitem[Celletti \& Giorgilli(1990)]{Celletti1990} Celletti, A. and Giorgilli, A.: On the stability of the Lagrangian points in the spatial restricted problem of three bodies. Celestial Mechanics and Dynamical Astronomy {\bf 50} (1), 31 (1990).

\bibitem[Chandrasekhar \& Contopoulos(1967)]{Contopoulos1967} Chandrasekhar, S. and G. Contopoulos.: On a post-Galilean transformation appropriate to the post-New\-to\-nian theory of Einstein, Infeld and Hoffmann. Proceedings of the Royal Society of London A: Mathematical, Physical and Engineering Sciences. {\bf 298} (1453), 123-141 (1967).

\bibitem[Chen \& Wu(2016)]{Chen2016} Chen, R. and Wu, X.: A Note on the Equivalence of Post-Newtonian Lagrangian and Hamiltonian Formulations. Commun. Theor. Phys. {\bf 65} (3), 321 (2016).

\bibitem[Contopoulos(1976)]{Contopoulos1976} Contopoulos, G. In Memoriam D. Eginitis, 159. D. Kotsakis, Ed., Athens (1976).

\bibitem[Contopoulos(2013)]{Contopoulos2013} Contopoulos, G. Order and chaos in dynamical astronomy. Springer Science \& Business Media (2013).

\bibitem[Damour et al.(2001)]{Damour2001} Damour, T., Jaranowski, P. and Schäfer, G.: Equivalence between the ADM-Hamiltonian and the harmonic-coordinates approaches to the third post-Newtonian dynamics of compact binaries. Physical Review D {\bf 63} (4), 044021 (2001).

\bibitem[Damour et al.(2002)]{Damour2002} Damour, T., Jaranowski, P. and Schäfer, G.: Erratum: Equivalence between the ADM-Hamiltonian and the harmonic - coordinates approaches to the third post-Newtonian dynamics of compact binaries. Physical Review D {\bf 66} (2), 029901 (2002).

\bibitem[de Andrade et al.(2001)]{Andrade2001} de Andrade, V.C., Blanchet, L. and Faye, G.: Third post-Newtonian dynamics of compact binaries: Noetherian conserved quantities and equivalence between the harmonic-coordinate and ADM-Hamiltonian formalisms. Classical and Quantum Gravity, {\bf 18} (5), 753 (2001).

\bibitem[Dubeibe \& Berm\'udez(2014)]{Dubeibe2014} Dubeibe, F. L. and Berm\'udez-Almanza, L. D.: Optimal conditions for the numerical calculation of the largest Lyapunov exponent for systems of ordinary differential equations. International Journal of Modern Physics C {\bf 25} (7), 1450024 (2014).

\bibitem[Eddington \& Clark(1938)]{Eddington1938} Eddington, A. and Clark, G. L.: The problem of n bodies in general relativity theory. In Proceedings of the Royal Society of London A: Mathematical, Physical and Engineering Sciences, {\bf 166} (927), 465-475 (1938).

\bibitem[Einstein et al.(1938)]{Einstein1938} Einstein, A., Infeld, L. and Hoffmann, B.: The gravitational equations and the problem of motion. Annals of Mathematics, 65-100 (1938).

\bibitem[Euler(1767)]{Euler67} Euler, L. Novo Comm. Acad. Sci. Imp. Petrop., 11, 144 (1767).

\bibitem[Froeschl\'e(1970)]{Froeschle1970} Froeschl\'e, C.: Numerical study of dynamical systems with three degrees of freedom. II. Numerical displays of four-dimensional sections. Astronomy and Astrophysics {\bf 5}, 177 (1970).

\bibitem[Gonczi(1981)]{Gonczi1981} Gonczi, R.: The Lyapunov characteristic exponents as indicators of stochasticity in the restricted three-body problem. Celestial mechanics {\bf 25} (3), 271-280 (1981).

\bibitem[Henon(1997)]{Henon-book} Henon, M. Generating families in the restricted three-body problem,  Springer-Verlag, Berlin (1997).

\bibitem[Huang \& Wu(2014)]{Huang2014} Huang, G. and Wu, X.: Dynamics of the post-Newtonian circular restricted three-body problem with compact objects. Physical Review D {\bf 89} (12), 124034 (2014).

\bibitem[Huang et al.(2016)]{Huang2016} Huang, L., Wu, X. and Ma, D.: Second post-Newtonian Lagrangian dynamics of spinning compact binaries. The European Physical Journal C {\bf 76} (9), 488 (2016).

\bibitem[Kla\v{c}ka \& Kocifaj(2008)]{Klacka2008} Kla\v{c}ka, J., and M. Kocifaj.: Times of inspiralling for interplanetary dust grains. Monthly Notices of the Royal Astronomical Society {\bf 390} (4), 1491 (2008).

\bibitem[Krefetz(1967)]{Krefetz1967} Krefetz, E.: Restricted three-body problem in the post-Newtonian approximation. The Astronomical Journal {\bf 72}, 471 (1967).

\bibitem[Landau(2013)]{Landau2013} Landau, L. D., The classical theory of fields. Vol. 2. ed. Elsevier  (2013).

\bibitem[Levi \& Steinhoff(2014)]{Levi2014} Levi, M. and Steinhoff, J.: Equivalence of ADM Hamiltonian and Effective Field Theory approaches at fourth post-Newtonian order for binary inspirals with spins. J. Cosmol. Astropart. Phys. {\bf12} (arXiv: 1408.5762), 003 (2014).


\bibitem[Lukes-Gerakopoulos(2014)]{Lukes2014} Lukes-Gerakopoulos, G. Adjusting chaotic indicators to curved spacetimes. Physical Review D {\bf 89} (4), 043002 (2014).

\bibitem[Maindl \& Dvorak(1994)]{Maindl1994} Maindl, T. I., and Dvorak, R.: On the dynamics of the relativistic restricted three-body problem. Astronomy and Astrophysics {\bf 290}, 335-339 (1994).

\bibitem[Marchal(2012)]{Marchal2012} Marchal, C., The three-body problem. Vol. 4. ed. Elsevier, (2012).

\bibitem[Motter \& Saa(2009)]{Motter2009} Motter, A. E., and Saa, A.: Relativistic invariance of Lyapunov exponents in bounded and unbounded systems. Physical review letters {\bf 102} (18), 184101 (2009).

\bibitem[Musielak \& Quarles(2014)]{Musielak2014} Musielak, Z. E., and B. Quarles.: The three-body problem. Reports on Progress in Physics, {\bf 77} (6) 065901 (2014).

\bibitem[Salazar et al.(2012)]{Salazar2012} Salazar, F. J. T., de Melo, C. F., Macau, E. E. N., and Winter, O. C.: Three-body problem, its Lagrangian points and how to exploit them using an alternative transfer to L4 and L5. Celestial Mechanics and Dynamical Astronomy, {\bf 114} (1-2), 201 (2012).

\bibitem[Su et al.(2016)]{Su2016} Su, XN., Wu, X. and Liu, FY.: Application of the logarithmic Hamiltonian algorithm to the circular restricted three-body problem with some post-Newtonian terms. Astrophysics and Space Science {\bf 361}, 32 (2016). 

\bibitem[Szebehely(1967)]{Szebehely1967} Szebehely V., Theory of orbits: the restricted problem of three bodies, Academic Press, New York (1967).

\bibitem[Tancredi et al.(2001)]{Tancredi2001} Tancredi, G., A. S\'anchez, and F. Roig.: A comparison between methods to compute Lyapunov exponents. The Astronomical Journal {\bf121}, (2), 1171 (2001).

\bibitem[Wang \& Huang(2015)]{Wang2015} Wang, H. and Huang, G.Q.: The Effect of Spin-Orbit Coupling and Spin-Spin Coupling of Compact Binaries on Chaos. Commun. Theor. Phys. {\bf 64} (2), 159 (2015).

\bibitem[Wu \& Huang(2003)]{Wu2003} Wu, X., and Huang, T.: Computation of Lyapunov exponents in general relativity. Physics Letters A {\bf 313}(1), 77-81  (2003).

\bibitem[Wu et al.(2006)]{Wu2006} Wu, X., Huang, T. and  Zhang, H.: Lyapunov indices with two nearby trajectories in a curved spacetime. Physical Review D {\bf 74} (8), 083001 (2006).

\bibitem[Wu et al.(2015)]{WuPRD} Wu, X., Mei, L., Huang, G. and Liu, S.: Analytical and numerical studies on differences between Lagrangian and Hamiltonian approaches at the same post-Newtonian order. Physical Review D {\bf 91}, 024042 (2015).

\bibitem[Wu \& Huang(2015)]{WuMNRAS} Wu, X. and Huang, G.: Ruling out chaos in comparable mass compact binary systems with one body spinning. Mon. Not. R. Astron. Soc. {\bf 452}, 3617 (2015).

\end{thebibliography}

\end{document}